\documentclass[aps,pre,twocolumn,superscriptaddress]{revtex4-1}

\usepackage[english]{babel}
\usepackage[utf8x]{inputenc}
\usepackage{amsmath,amsfonts,amssymb,amsbsy,eucal,dsfont}
\usepackage{color}
\usepackage[colorlinks]{hyperref}
\usepackage{graphicx}
\usepackage{rotating}

\newcommand{\hide}[1]{}

\begin{document}

\title{Viscosity and Diffusion: Crowding and Salt Effects in Protein Solutions}

\author{Marco Heinen}
\affiliation{Institute of Complex Systems (ICS-3), Forschungszentrum J\"ulich, D-52425 J\"ulich, Germany}

\author{Fabio Zanini}
\affiliation{Institut f\"ur Angewandte Physik, Universit\"at T\"ubingen, 72076 T\"ubingen, Germany}

\author{Felix Roosen-Runge}
\affiliation{Institut f\"ur Angewandte Physik, Universit\"at T\"ubingen, 72076 T\"ubingen, Germany}

\author{Diana Fedunov\'a}
\affiliation{Institute of Experimental Physics, Slovak Academy of Sciences, 04353 Ko\v{s}ice, Slovakia}

\author{Fajun Zhang}
\affiliation{Institut f\"ur Angewandte Physik, Universit\"at T\"ubingen, 72076 T\"ubingen, Germany}

\author{Marcus Hennig}
\affiliation{Institut f\"ur Angewandte Physik, Universit\"at T\"ubingen, 72076 T\"ubingen, Germany}
\affiliation{Institut Laue-Langevin, 38042 Grenoble Cedex 9, France}

\author{Tilo Seydel}
\affiliation{Institut Laue-Langevin, 38042 Grenoble Cedex 9, France}

\author{Ralf Schweins}
\affiliation{Institut Laue-Langevin, 38042 Grenoble Cedex 9, France}

\author{Michael Sztucki}
\affiliation{European Synchrotron Radiation Facility, 38043 Grenoble Cedex 9, France}

\author{Mari\'an Antal\'ik}
\affiliation{Institute of Experimental Physics, Slovak Academy of Sciences, 04353 Ko\v{s}ice, Slovakia}
\affiliation{Department of Biochemistry, Faculty of Sciences, P. J. \v{S}af\'{a}rik University, Moyzesova 11, 04154 Ko\v{s}ice, Slovakia}

\author{Frank Schreiber}
\email[]{frank.schreiber@uni-tuebingen.de}
\affiliation{Institut f\"ur Angewandte Physik, Universit\"at T\"ubingen, 72076 T\"ubingen, Germany}

\author{Gerhard N\"agele}
\email[]{g.naegele@fz-juelich.de}
\affiliation{Institute of Complex Systems (ICS-3), Forschungszentrum J\"ulich, D-52425 J\"ulich, Germany}

\date{\today}

\renewcommand{\figurename}{Fig.}
\renewcommand{\figuresname}{Figs.}
\renewcommand{\refname}{Ref.}
\newcommand{\refsname}{Refs.}
\newcommand{\expressionname}{Eq.}
\newcommand{\expressionsname}{Eqs.}
\newcommand{\sectionname}{Sec.}

\begin{abstract}
We report on a joint experimental-theoretical study of collective diffusion in, and static shear viscosity of solutions of 
bovine serum albumin (BSA) proteins, focusing on the dependence on protein and salt concentration. Data obtained from dynamic
light scattering and rheometric measurements are compared to theoretical calculations based on an analytically treatable
spheroid model of BSA with isotropic screened Coulomb plus hard-sphere interactions.
The only input to the dynamics calculations is the static structure factor obtained from a consistent theoretical fit to a concentration series of
small-angle X-ray scattering (SAXS) data. This fit is based on an integral equation scheme that combines high accuracy with low computational cost.
All experimentally probed dynamic and static properties are reproduced theoretically with an at least semi-quantitative
accuracy. For lower protein concentration and low salinity, both theory and experiment show a maximum in the reduced viscosity, caused by
the electrostatic repulsion of proteins. The validity range of a generalized Stokes-Einstein (GSE) relation connecting viscosity, collective diffusion coefficient,
and osmotic compressibility, proposed by Kholodenko and Douglas [PRE, 1995, \textbf{51}, 1081] is examined.
Significant violation of the GSE relation is found, both in experimental data and in theoretical models, in semi-dilute
systems at physiological salinity, and under low-salt conditions for arbitrary protein concentrations.
\end{abstract}

\maketitle

\section{Introduction}
A quantitative understanding of the dynamics in concentrated solutions of interacting proteins is of importance to the evaluation
of cellular functions, and the improvement of drug delivery. Transport properties such as collective and self-diffusion coefficients,
and the static and high-frequency shear viscosities, are strongly affected by the aqueous environment~\cite{Ball2008}, and
in particular by crowding effects due to high concentration of macromolecules,
coupled both by direct and solvent-mediated, hydrodynamic interactions (HIs)~\cite{Ellis2001, Zimmerman1993, Skolnick2010}.
The latter type of interaction, which is both long-ranged and of many-body nature, poses a particularly challenging task
to a theoretical treatment of diffusion and rheological transport properties. 

In the present paper, we report on a combined experimental and theoretical study on collective diffusion,
low shear-rate static viscosity, and static and dynamic scattering functions of concentrated solutions of bovine serum albumin (BSA) proteins.
The goal of this study is twofold. On the one hand, we explore how far a simple colloidal model in combination with state-of-the-art theoretical schemes
can capture the microstructure and dynamics of proteins in solution. On the other hand, we investigate the
concentration- and salt-dependence of collective diffusion and the static shear viscosity, and use our results to test the  
validity range of a generalized Stokes-Einstein (GSE) relation which combines the collective diffusion coefficient with the isothermal osmotic
compressibility and the shear viscosity.

BSA is a protein which is readily soluble in water and stable over a wide range of salt and protein concentrations.
Its stability and reproducibility make it well-suited as a model system of globular proteins.
Proteins constitute identical solute units surpassing any synthetic colloid suspension in terms of monodispersity.
In this respect, they are ideally suited to the application of analytical theoretical models used with good success for
large colloids. However, the construction of a quantitatively accurate theoretical model for protein solutions
is considerably obstructed not only by the potential presence of impurities and oligomers,
but also by the complex internal conformation and surface of a protein.
The folding state depends on various control parameters such as temperature, protein concentration, pH value, and salinity.
The irregular protein surface
implies an orientation-dependent protein interaction 
energy with repulsive and attractive parts, and furthermore complicates the description of hydroynamically influenced transport properties.
 
In a first step towards calculating dynamic properties of proteins, it is nonetheless possible to use a model of reduced complexity,
with system parameters such as the pH-dependent particle charge determined from a consistent fit of theoretical expressions for the scattered intensity
to the experimental static scattering functions. We use here a simple colloid model where the BSA interactions are described by the repulsive,
electrostatic plus hard-core part of the isotropic Derjaguin-Landau-Verwey-Overbeek (DLVO) potential \cite{Verwey_Overbeek1948}. 
The effect of the non-spherical shape of BSA proteins is accounted for in the static intensity calculations within the so-called 
translational-orientational decoupling approximation, by describing the proteins as oblate spheroids interacting by a spherically symmetric
effective pair potential.

Using this simplifying protein interaction model, the static structure factor, $S(q)$, entering into the static scattered intensity,
is calculated as a function of wavenumber $q$, by using our newly developed modified penetrating background corrected rescaled mean spherical approximation
(MPB-RMSA). This analytical method has been shown to be in excellent accord with numerically expensive computer simulation results for $S(q)$
\cite{Heinen2011, Heinen_Erratum_2011}. The system parameters of the protein-interaction model, most notably the effective protein charge,
are determined from adjusting the theoretically calculated static intensity, $I(q)$,
to the experimental one. The consistent agreement of calculated values and small-angle X-ray scattering (SAXS)
data for $I(q)$ in a wide range of concentrations and wavenumbers indicates that left-out attractive interaction contributions are of minor
importance at the considered salinities. As an independent additional check, the static light scattering (SLS) data for
$S(q)$ at low $q$ are found to be well reproduced by the theoretical fits of the SAXS data.

Without any further adjustment, the analytically calculated static structure factors are used as the only input to our theoretical
calculations of the collective diffusion coefficient, $d_C$, and the low shear-rate limiting static viscosity $\eta$. To calculate $d_C$ and the
high-frequency part, $\eta_\infty$, of the static viscosity, we use two approximate analytical schemes, namely the pairwise additive hydrodynamic interaction
(PA) approximation, and the so-called self-part corrected $\delta\gamma$ method. As shown by two of the present authors \cite{Heinen_TheoArticlePreparation},
these two methods give results which are in general in good agreement with more elaborate Stokesian Dynamics simulation results for particles with Yukawa-type pair interactions. 

The static viscosity,
\begin{equation}\label{eq:eta_stat}
\eta = \eta_\infty + \Delta\eta, 
\end{equation}
consists of a short-time part, $\eta_\infty$, determined solely by hydrodynamic interactions (HIs), and a shear-stress relaxation part $\Delta\eta$, with $\Delta\eta > 0$.
We calculate the latter using mode-coupling theory (MCT), which, like the two employed short-time schemes, requires $S(q)$ as the only input. 

Our comparison with the experimental $d_C$ measured by dynamic light scattering (DLS), and with $\eta$ obtained from viscometry,
is a stringent test for our theoretical results and for the employed isotropic interaction model, since except for the static input,
no fit parameters are involved. In particular, no further adjustments of the theoretical predictions have been made on referring
to the actually non-spherical shape of BSA proteins. We show that despite the simplicity of our model, most dynamic features are 
well reproduced by the theoretical results, to an at least semi-quantitative accuracy. In particular, both a low-concentration maximum of the
reduced viscosity, and a maximum in $d_C$ at a different concentration, are well captured by the theory.

For BSA, also the short-time self-diffusion has been recently found to be reasonably well described by a simple spheroid model~\cite{Roosen-Runge2011,Roosen-Runge2010}.
Of course, this does not imply that the complex conformation of a globular protein plays no role.
The DLVO model (even with inclusion of van der Waals attraction) is not sufficient to fully explain the rich phase behavior of proteins.
For example, it has been shown that surface patchiness has an important effect on the phase diagram~\cite{Goegelein2008}.
Also, binding of multivalent ions to the protein surface can give rise to non mean-field behaviors beyond DLVO, such as charge inversion,
re-entrant condensation and liquid-liquid phase separation~\cite{Zhang2008,Zhang2010,Zhang2011}.

Generalized Stokes-Einstein (GSE) relations, which approximately relate diffusion to rheological properties in concentrated complex liquids, are 
an important issue in microrheological studies, since a valid GSE relation allows to infer a rheological property more easily from diffusion measurements.
Several GSE relations in colloidal dispersions of electrically neutral (porous and non-porous) spheres, and charged particle suspensions have been
explored~\cite{Banchio1999GSESs, Bergenholtz1998GSE, Banchio2008, Abade2010GSE}.
We study here a GSE relation not discussed in this earlier work, which has been
proposed by Kholodenko and Douglas~\cite{Kholodenko1995}. This GSE relation, which we refer to in the following as the KD-GSE relation,
has been used in the biophysical and soft matter community \cite{Gaigalas1995, Cohen1998, Boogerd2001, Nettesheim2008}.
It relates $d_C$ to $\eta$, and to the square-root of the isothermal osmotic compressibility.

We present a thorough discussion of the validity range
of the KD-GSE relation for BSA solutions, and for generic colloidal fluids of particles with screened Coulomb interactions, for a large range of salinities.
Both the short-time and the long-time versions of the KD-GSE relation are considered. At high salinity, where the electrostatic interaction of particles is strongly screened,
we find these two relations to become invalid at larger concentrations. At lower salinity, the KD-GSE relations are poorly satisfied even at low concentrations.  
   
The paper is organized as follows. Sec.~\ref{sec:Exp_Details} includes the experimental details of the sample preparation, and of the 
SLS, DLS, SAXS, and rheological measurements. In Sec.~\ref{sec:Theory}, we discuss the employed simplifying model of BSA, and present the essentials
of our theoretical methods, allowing for a fast calculation of measured static and dynamic properties.
Our experiment data are shown in combination with the theoretical results in 
Sections \ref{sec:Static_Exp_vs_Theo} and \ref{sec:Dynamic_Exp_vs_Theo}, dealing with static and dynamic properties, respectively. Sec.~\ref{sec:Dynamic_Exp_vs_Theo}
includes the examination of the KD-GSE relation.
Our conclusions are contained in Sec.~\ref{sec:conclusions}.

\section{Experimental details}\label{sec:Exp_Details}

\subsection*{Sample preparation}

BSA is a globular protein with a linear extension of about $7$ nm.
The considered aqueous solutions of BSA with no added salt, and with monovalent added salt such as NaCl, have a pH in between 5.5 and 7.
Under these conditions, BSA is stable in solution, folded in its native state, and carrying a negative 
net charge in the range of roughly $8$ to $20$ elementary charge units (see below for details)~\cite{peters1985serum,Bohme2007}.
BSA was purchased from Sigma (cat. A3059) as a lyophilized powder, certified globulin- and protease free.

The sample preparation for all experimental techniques started with the dissolution of protein powder in a solvent, and subsequent waiting
until the solution was homogenized. The protein mass concentration, $c_p$, in the solution volume, $V_{\textrm{H}_2\textrm{O}} + m_p \cdot \theta$,
is given by the BSA weight $m_p$ via
\begin{equation}\label{eq:cp_exp}
 c_p = \frac{m_p}{V_{\textrm{H}_2\textrm{O}} + m_p \cdot \theta} \quad ,
\end{equation}
with the specific protein volume $\theta = 0.74$ ml/g~\cite{Lee1974} determines the self-volume of proteins upon dissolution.

For small-angle X-ray scattering, deionized and degased water was used as solvent.
The samples with concentrations higher than 15 mg/ml were prepared directly, while smaller concentrations were prepared from a stock solution of 18 mg/ml.
The samples were filled into a plastic syringe and inserted into the capillary during the measurement.

For the viscosity measurements, the solutions were prepared similarly using as solvent both deionized water, and solutions of NaCl in deionized water.
The NaCl molarity is calculated from the total solution volume, including the protein self-volume.
All solutions used for the viscosity experiments were further degased by a water-jet air-pump.

For our light scattering experiments, stock solutions of BSA proteins in deionized water were mixed with solutions of NaCl in deionized
water according to the required concentration. The NaCl molarity is calculated from the total water volume.
Then, every sample was pressed with a plastic syringe through a hydrophilized nylon membrane filter with a pore size of $100$ nm (Whatman Puradisc 13),
and transferred into a cylindrical glass scattering cell.
The cell was sealed immediately with a plastic cap.

The effect of the difference in NaCl concentrations between light scattering and viscosity samples, arising from the slightly
differing sample preparation, is negligibly small.

\subsection*{Static and dynamic light scattering}

Multi-angle dynamic light scattering (DLS) was performed at various concentrations of protein and added salt, at a temperature of $T =295~\text{K}$.
In particular, the BSA mass concentration, $c_p$, was chosen between $0.1$ to $150$ mg/ml, and the concentration of added salt
was $0$ (no added salt), $5$, $150$ and $500~\text{mM}$.
Note that, even in the zero added-salt case, the analysis of the scattering data discussed in Sec.~\ref{sec:Static_Exp_vs_Theo} reveals a residual
electrolyte concentration of a few mM, scaling roughly linear with $c_p$ (see \tablename~\ref{tab:lowsalt_fitparams}).
This suggests a few possible sources of the residual electrolyte ions. Firstly, a possible source could be the surface-released counterions
of charged BSA oligomers, not contained in our 
monodisperse model. Secondly, a salt contamination of the BSA stock, and thirdly the dissociation of 
acidic or alkaline surface groups off the BSA proteins cannot be excluded.
   
Static light scattering (SLS) experiments were performed on the same samples.
We used a combined SLS/DLS device from ALV (goniometer: CGS3, correlator: 7004/FAST), located at the Institut Laue Langevin in Grenoble,
with a minimum correlation time of $3.125~\text{ns}$ as initial and shortest time. The HeNe laser was operating at wavelength $\lambda_0 = 632.8~\text{nm}$,
with an output power of $22~\text{mW}$. The accessible range for the scattering angle (wavenumber) was
$30~\text{-}~150^\circ$ (q = $0.007~\text{-}~0.026~\text{nm}^{-1}$).
Moreover, the DLS intensity autocorrelation
function decays on a time scale much slower than the interaction time, $\tau_I \sim \sigma_{d_0}^2/(4 d_0) \sim 0.3~\mu\text{s}$, of BSA, where $d_0$ is the 
single protein average translational free-diffusion coefficient, and $\sigma_{d_0}$ is an effective hydrodynamic diameter.
Hence, DLS probes the long-time collective diffusion of BSA, in the $q\to 0$ limit.

The normalized intensity autocorrelation function obtained from DLS,
\begin{equation}
g_2(q,t) = \dfrac{\left<I(\mathbf{q},0) I(\mathbf{q},t) \right>}{{\left<I(\mathbf{q})\right>}^2},\nonumber 
\end{equation}
was fitted according to the Siegert relation, by the double exponential decay function
\begin{equation}\label{eq:Double_expo}
g_2(q,t) - 1 = \left( \sum_{i=1,2} A_i \cdot \exp\left[ - D_i \, q^2 \, t  \right] \right)^2 ~  + \textrm{B},
\end{equation}
with decay constants $D_1$ and $D_2$, and amplitudes $A_1$ and $A_2$. 
The fit results were essentially the same with and without the background-correction constant $\textrm{B}$.
At all probed angles, the two decay constants are widely separated ($D_1 \gg D_2$). The faster mode, $D_1$, is attributed to the
(long-time) collective diffusion coefficient, $d_C$, of BSA monomers. The appearance of the slower mode characterized by $D_2$, can be attributed to
the slow motion of the larger impurities and oligomers.
After having checked that $D_1$ is overall $q$-independent within the experimental resolution, it was averaged with respect to its residual scattering angle fluctuations
to gain better statistics.
Data on $D_2$ are rather noisy in comparison to $D_1$, and show no clear dependence on $q$, $c_p$, and on the concentration of added salt.

\subsection*{Small-angle X-ray scattering}

Aqueous solutions of BSA with mass concentrations between $0.9~\text{mg/ml}$ and $270~\text{mg/ml}$, and  without added salt, were measured
by small-angle X-ray scattering (SAXS), at the beam line ID02 of the European Synchrotron Radiation Facility (ESRF) in Grenoble, France.
The standard configuration at a 2m sample-to-detector distance, and a photon energy of 16051 eV was used. Measurements were repeated several
times in the flow mode and with short detection times to ensure the absence of radiation damage.
The data from the CCD were processed with the standard routines available at the beam line for radially averaging the data and correcting
for transmission. Repeated measurements were summed up, and the solvent scattering was measured independently and subtracted from the data.
Additionally, two dilute samples (at $c_p = 1$ and $2$ mg/ml) with $150~\text{mM}$ of added NaCl were measured for form factor fitting.

\subsection*{Viscosity measurements}

The viscosity data were measured at $T =25^\circ$ C, for different concentrations of protein and added salt.
The first dataset was obtained for solutions without added salt, while the second set describes systems with $150~\text{mM}$ NaCl.
All measurements were performed at a shear rate of $60~\text{Hz} \ll 1/\tau_I$,
using the suspended couette-type viscometer described in Ref.~\cite{Bano2003}. The important advantage of this instrument is the
possibility to collect data without errors caused by the surface shear-viscosity. A test made for $c_p \approx 20$ mg/ml and $100$ mg/ml, 
without salt and for 2M added NaCl, revealed no shear-rate dependence of the viscosity for shear rates between $50$ and $95~\text{Hz}$.
The precision of the viscosity measurements is approximately $0.1\%$. In order to minimize systematic errors,
every measurement was repeated three times, including separate sample preparations.

The viscometer directly measures the relative shear-viscosity of the solution against pure water
(for technical details see Ref.~\cite{Bano2003}). For the aqueous BSA solutions without added salt discussed in this work, the relative viscosity
was directly measured. For BSA solutions with added salt, this quantity was obtained as the ratio of the following two values:
(a) the directly measured relative viscosity of the BSA solution with salt against water divided by (b) the directly measured relative viscosity of
the salt solution (without BSA) against water.

\section{Theory}\label{sec:Theory}

\subsection*{Single-particle properties}

In the following, we discuss the spheroid model of BSA. We use this model for the form factor fitting, and in determining effective sphere
diameters related to different single-particle properties.

At low protein concentration and sufficient amount of added salt, inter-protein correlations are negligible. The scattered intensity, $I(q)$, is then solely
determined by the form factor $P$, i.e. $I(q) \propto P(q)$. Crystallographic measurements \cite{Carter1994, Ferrer2001,Leggio2008} have revealed a flat
and roughly heart-shaped structure of albumins. The computation of
single-particle properties with an account of the highly complex particle shape of biomolecules can be done by numerical simulations only and is beyond the scope of
this paper \cite{DeLaTorre2010, Ferrer2001}. Rather, the aim of the present study is to give an essentially analytic description of the microstructure
and the dynamics of interacting BSA proteins with low computational cost.
We therefore intentionally choose an extremely simple model for the fit of the protein form factor, by an oblate, solid ellipsoid (spheroid).
Clearly, this mapping of the complex protein configuration onto an essential geometric shape is a delicate and broad topic on its own.
Considering that the focus of the present work is on collective correlations rather than on single-particle properties, we cannot
discuss all details of this subtle matter; we basically follow the approach of Ref. \cite{Zhang2007}.

For a homogeneously scattering spheroid with dimensions $a$ and $b$, where $a$ denotes the semi-axis of revolution, the orientationally
averaged form factor, $P_{ell}$, is given by \cite{Pedersen1997}
\begin{equation}
P_{ell}(q) = \int_0^1 d\mu {|f(q,\mu)|}^2\label{eq:Ellipse_Formfact}
\end{equation}
with the scattering amplitude $f(q,\mu) =3 j_1(u)/u$, and $u =q\sqrt{a^2\mu^2 + b^2(1-\mu^2)}$. Here, $j_1$ is the spherical Bessel function of the first kind.

The fit of Eq.~\eqref{eq:Ellipse_Formfact} to our newly recorded, low-concentration SAXS intensities at $c_p =1$ and $2$~mg/ml, and for $150$~mM of added NaCl, is shown in
\figurename~\ref{fig:Formfakt_fit} of Sec.~\ref{sec:Static_Exp_vs_Theo}, along with a discussion of the obtained best fit values $a =1.75$~nm and $b =4.74$~nm.

When protein correlations come into play at higher concentrations or lower salinities, the spheroid model of BSA becomes too complex for an analytic treatment.
Therefore, as far as the protein-protein interactions are concerned, we describe the proteins as effective spheres with diameter $\sigma$.
Depending on the considered single-particle property, different definitions for $\sigma$ can be given.

Consider first the geometric effective diameter, $\sigma_{geo} = 8(ab^2)^{1/3} = 6.80~\text{nm}$, which follows from equating the volume of the 
effective sphere to that of the spheroid.
This effective diameter reflects the volume of the protein and the hydration layer visible to SAXS, but does not include thermo- and hydrodynamic
effects of non-sphericity~\cite{Svergun1998,Zhang2011a}. Thus, it should be considered as a lower boundary to the effective sphere diameter.

A thermodynamic effective diameter, $\sigma_{B_2} = 7.40~\text{nm}$, follows from demanding equal second virial coefficients, $B_2(T)$, of hard spheroid and
effective hard sphere \cite{Isihara1950}.

Alternatively, dynamic single-particle properties can be used in defining the effective diameter. For hydrodynamic stick-boundary conditions and $a<b$,
the translational free diffusion coefficient of an isolated spheroid reads \cite{Jennings1988,Ferrer2001,Perrin1936}
\begin{equation}
d_0^{ell}(a,b) = \dfrac{k_B T S(a,b)}{12 \pi \eta_0 a},\label{eq:d0_spheroid}\\ 
\end{equation}
with absolute temperature $T$, Boltzmann's constant $k_B$, solvent shear-viscosity $\eta_0$,
$S(a,b) = 2~\text{atan}~\xi(a,b)/\xi(a,b)$ and $\xi(a,b) = \sqrt{\left|a^2-b^2\right|}~/a$. Equating $d_0^{ell}$ to
the diffusion coefficient, $d_0 = k_B T / (3 \pi \eta_0 \sigma_{d_0})$, of an effective sphere gives $\sigma_{d_0} = 7.38~\text{nm}$.

Finally, one can derive another effective diameter from the intrinsic viscosity
\begin{equation}\label{eq:eta_intrins_definition}
[\eta] = \lim_{\phi\to0}\frac{\eta(\phi)-\eta_0}{\eta_0 \phi}, 
\end{equation}
where $\phi$ is the particle volume fraction. 
For a spheroid with hydrodynamic stick-boundary conditions \cite{Jeffery1922, Philipse2001},
\begin{equation}\label{eq:eta_intrins_spheroids}
[\eta]^{ell} = \dfrac{5}{2} + \frac{32}{15\pi}\left[\dfrac{b}{a}-1\right] - 0.628 \left[ \frac{1-a/b}{1-0.075 a/b} \right],
\end{equation}
which for $a = b$ reduces to the Einstein result, $[\eta]^{sph} = 2.5$, for a solid sphere.
Note here that $[\eta]^{ell} > 2.5$ for $a \neq b$. Explicitly, $[\eta]^{ell} = 3.25$ for the best fit values $a$ and $b$ given in \figurename~\ref{fig:Formfakt_fit}.
On demanding equality of the interaction-independent linear terms in the virial expansions of the viscosity,
\begin{displaymath}
\dfrac{\eta}{\eta_0} = 1 + [\eta]\phi + \mathcal{O}(\phi^2),
\end{displaymath}
for spheroids and effective spheres, and on using $\phi^{ell} = (4\pi/3) a b^2 n$ and
$\phi^{sph} = (\pi/6) {\sigma_{[\eta]}}^3 n$ for an equal number density $n$, the effective diameter $\sigma_{[\eta]} = 7.42~\text{nm}$ is obtained.

Since the aspect ratio, $b/a = 2.71$, is rather close to unity, the four obtained effective diameters are quite similar in magnitude.
We use $\sigma = \sigma_{B_2} = 7.40~\text{nm}$ in all our calculations of static and dynamic properties discussed in this paper.

\subsection*{Static scattering intensity and structure factor}

Concentrated protein solutions exhibit pronounced inter-particle correlations which are reflected in the static scattering intensity. This 
applies also to dilute, low-salinity solutions where the proteins show long-ranged electrostatic repulsion. 

In order to allow for an analytical theoretical treatment, we assume that the static scattering intensity of interacting BSA proteins can be approximated by
\begin{equation}\label{eq:Iq_decoupling}
I(q) = A c_p P_{ell}(q)S_m(q),
\end{equation}
where $S_m$ is the so-called measurable static structure factor.
Here, $A$ is a $q$-independent factor (of dimension $\text{velocity}^3$),
that should be the same for all intensity measurements corrected for recording time and source intensity.

For calculating $S_m(q)$, we use the rotational-translational decoupling approximation \cite{Nagele1996,Kotlarchyk1983},
where the spheroid shape is accounted for in the scattering
amplitudes only, so that 
\begin{equation}
S_m(q) =  \left[\rule[-.5em]{0em}{1.5em}1-X(q)\right] + X(q)S(q). \label{eq:Decoupling_S}
\end{equation}
Here,
\begin{equation}
X(q) = \dfrac{1}{P_{ell}(q)} {\left[\int\limits_0^1 d\mu f(q,\mu) \right]}^2,
\end{equation}
with $0\leq X(q) \leq 1$ and $X(q\to 0) =1$, and $S$ is the so-called ideal structure factor of ideally monodisperse effective spheres of diameter $\sigma = \sigma_{B_2}$
and screened Coulomb repulsion of DLVO type.
For the BSA model spheroid used here, $X(q)$ stays close to unity for
$q \lesssim 0.5~\text{nm}^{-1}$, decaying for larger $q$ steeply towards its first zero value at $q \approx 1.3~\text{nm}^{-1}$. For $q>1.3~\text{nm}^{-1}$, $X(q) < 0.04$.
The orientational disorder assumed in the decoupling approximation has the general effect of damping the oscillations in $S_m(q)$. 
While $S_m(q)$ is practically equal to one for $q \gtrsim 1.3~\text{nm}^{-1}$, irrespective of the still visible oscillations in $S(q)$, 
the effect of orientational disorder on $S_m(q)$ is weak in the range $q \lesssim 0.5~\text{nm}^{-1}$, where the most distinctive features in $S(q)$ are seen.
We further note that $S_m(q\to0) = S(q\to0)$ for monodisperse systems, a feature which plays an important role in our upcoming discussion of collective diffusion.

The ideal structure factor, $S(q)$, entering into Eq.~\eqref{eq:Decoupling_S}, is calculated using the repulsive part of the
DLVO pair-potential \cite{Verwey_Overbeek1948},
\begin{equation}\label{eq:HSY_pair_pot}
 \beta u(x) = \left\lbrace
   \begin{array}{ll}
   \infty \,, & x = r/\sigma < 1,\\
   \gamma\;\! \dfrac{{e^{-k x}}}{x} \,, & x > 1,
   \end{array}
 \right.
\end{equation}
also referred to as the hard-sphere Yukawa (HSY) potential. The coupling parameter, $\gamma$, and the screening parameter, $k$, are given by
\begin{subequations}
\begin{eqnarray}
  \gamma &=& \frac{L_B}{\sigma}\left( \frac{e^{k/2}}{1+k/2} \right)^2 \! Z^2,\label{eq:DLVOcouplingConst}\\
   k^2 &=& k^2_c + k^2_s =  \frac{L_B / \sigma}{1 - \phi}\;\! \left( 24 \phi |Z| + 8 \pi n_s \sigma^3  \right). \label{eq:DLVOSscreeningConst}
\end{eqnarray}
\end{subequations}
Here, $L_B = \beta e^2/\epsilon$ is the solvent-characteristic Bjerrum length in Gaussian units, $\beta = 1/(k_B T)$,
$\epsilon$ is the solvent dielectric constant, and $Z$ is the effective protein charge number in units of the proton elementary
charge $e$. The factor $1/(1 - \phi)$ in $k^2$ corrects for the free volume available to the microions \cite{Russel1981,Denton2000}.
We have not included van der Waals (vdW) forces in $u(x)$. However, we have checked that the influence of vdW attractions
is small for most of the considered systems.  

Eq.~\eqref{eq:DLVOSscreeningConst} consists of two additive parts.
The first part, $k^2_c \propto |Z|$, is due to protein-surface released counterions, which are assumed to be monovalent. 
The second part, $k^2_s$, accounts for the screening due to all other monovalent microions. Owing to the overall charge neutrality, this contribution is 
proportional to the co-ion concentration $n_s$. 
A lower bound of $n_s \geq 10^{-7}$ M
in pH-neutral aqueous solutions is due to the self-dissociation of water.
Additional contributions to $n_s$ can arise from dissolved CO$_2$, and added salt
such as NaCl. For a protein solution, $n_s$ can have a (putatively linear)
dependence on $c_p$ if charged protein oligomers are present, acting as an additional source of surface-released counterions not
contained in our model. Moreover, the protein stock solution might contain a residual amount of salt, and the proteins might dissociate
acidic or alkaline surface groups during solvation. 
Note that due to the overall charge neutrality, the total concentration of monovalent counterions is given by
$n_s + 6\phi|Z|/(\pi \sigma^3)$.

In recent work \cite{Heinen2011, Heinen_Erratum_2011}, two of the present authors have derived a computationally efficient integral equation scheme
for computing $S(q)$ using the screened Coulomb potential in Eq.~\eqref{eq:HSY_pair_pot}.
This so-called modified penetrating background corrected rescaled mean spherical approximation (MPB-RMSA) shares the analytical simplicity
of the widely used RMSA \cite{Hansen1982,Zhang2007}, but is distinctly more accurate.
All calculations of $S(q)$ in this paper are based on the MPB-RMSA.  

The spheroid-Yukawa (SY) model used in our calculations of $I(q)$ and $S_m(q)$ ignores orientational-translational coupling. 
Therefore, it can be expected to apply only to fluid-phase BSA solutions when $c_p$ is sufficiently low, and 
when the ionic strength is not too large, so that the anisotropic protein shape and pair-interaction parts are not important.
At larger $c_p$, there is orientational-translational coupling, and the decoupling
approximation becomes invalid. We note again that the possible presence of residual protein oligomers and 
scattering impurities is not accounted for in our one-component model. The virtue of the SY model, however,
is its analytical simplicity. The concentration range in which the SY model is applicable to BSA is examined in Sec.~\ref{sec:Static_Exp_vs_Theo}.   

Since we use a spherically symmetric
screened Coulomb plus hard-core pair potential for the protein-protein interactions, a short discussion of the neglected
anisotropy in the electric double layer around a charged spheroid is in order here.

The mean electrostatic potential, $\Phi(r,\mu) = \sum_{l=0}^\infty \Phi_{l}(r) P_l(\mu)$, of a spheroid with a corresponding
axisymmetric charge distribution immersed in an electrolyte solution
includes in general higher-order multipoles with $l > 0$.  Here, $r$ is the distance of the spheroid center to the field point,
$\mu = \cos \vartheta$ is the cosine of the angle relative to the spheroid rotational symmetry axis, and the
$P_l$'s are Legendre polynomials.  

For large $r$, all multipoles decay asymptotically equally fast according to \cite{Yoon1991, vanRoij2009, vanRoij2011, Tellez2010, Kjellander2003, Likos2004}
\begin{equation}\label{eq:farfield_Spheroid_pot}
\Phi_{l}(r) \sim f_{l} \dfrac{e^{-\kappa r}}{r},
\end{equation}
where $\kappa$ denotes the inverse electrostatic screening length, and $f_{l}$ depends on the charge distribution.
This implies that, in principle, the pair interaction energy of two spheroids 
depends on their relative orientation even when $r \gg \kappa^{-1}$. However, the multipolar strengths, $f_l$, for a spheroid with $b/a \sim 1$
can be expected to be small for larger $l$. Moreover, since after orientational averaging, $\left< P_l(\mu) \right>_\mu = 0$ for all $l > 0$,
our neglect of anisotropic pair interaction contributions can be expected to be reasonable, for systems where the particles
can essentially rotate freely.

\subsection*{Short-time diffusion}

We summarize here the analytical methods used in calculating the (short-time) collective diffusion coefficient $d_C$.
These methods require $S(q)$ as their only input, with the BSA protein interactions described by the spherical pair potential in Eq.~\eqref{eq:HSY_pair_pot}.

The colloidal short-time regime covers correlation times $t$ within $\tau_B \ll t \ll \tau_I$. Here, $\tau_B = m_p/(3\pi\eta_0 \sigma)$
is the momentum relaxation time of a globular protein of mass $m_p$. Within a short-time span, a protein has diffused a very small
fraction of its size only. For BSA in water, $\tau_B \sim 1$ ps, and $\tau_I \approx0.3$ $\mu$s. The BSA short-time
dynamics is thus not resolved in our DLS experiment determining the measurable dynamic structure factor, $S_m(q,t)$, as a
function of wavenumber $q$ and correlation time $t$.  

Within the translational-orientational decoupling approximation used in the SY model,
$S_m(q,t)$ is determined by the right-hand-side of
Eq.~\eqref{eq:Decoupling_S} with $S(q)$ replaced by $S(q,t)$. The latter is the ideal dynamic structure factor of ideally monodisperse, charged effective
spheres interacting according to Eq.~\eqref{eq:HSY_pair_pot}.

Owing to the smallness of the proteins compared to the wavelength of visible laser light used in our DLS experiments, one obtains $t \gg \tau_I$ and $q \ll q_m$.
Here, $q_m$ is the wavenumber where $S$ attains its principle peak value. Since $X(q\ll q_m) \approx 1$, it follows that $S_m(q\ll q_m,t) \approx S(q\ll q_m,t)$, so that 
the influence of orientational disorder on the measured $S_m(q,t)$ via the spheroid form factor is negligible.     

As a consequence, DLS determines the long-time collective diffusion coefficient, $d_C^L$, according to
\begin{equation}\label{eq:S_qt_decay}
S_m(q \ll q_m, t \gg \tau_I) \propto \exp \left[ -q^2 d_C^L t \right].
\end{equation}
The coefficient $d_C^L$, also referred to as the gradient diffusion coefficient,
quantifies the long-time decay of long-wavelength, isothermal protein concentration fluctuations.
In Eq.~\eqref{eq:S_qt_decay}, additional scattering contributions to $S_m(q,t)$, originating from oligomers and large impurities, are neglected.
As discussed in relation to Eq.~\eqref{eq:Double_expo}, these give rise to an additional, exponentially decaying mode with a mean
diffusion constant, $D_2$, which is substantially smaller than $d_C^L$.  

While, in principle, $d_C^L$ needs to be distinguished from its short-time counterpart $d_C^S$, with $d_C^L \leq d_C^S$, it has
been shown \cite{Nagele1996, Szymczak2004} that the relative difference is very small ($\lesssim 5\%$) even in highly concentrated systems.
For solutions like the ones considered in this work, where non-pairwise additive HI contributions are small, $d_C = d_C^L$ becomes
practically identical to $d_C^S$. This allows us to use more simple short-time dynamic methods for calculating $d_C$. 

To this end, we use two complementary analytical methods, namely a self-part corrected version of the so-called $\delta\gamma$ scheme due
to Beenakker and Mazur \cite{Heinen_TheoArticlePreparation, heinen2010short, Beenakker1983, Beenakker1984, Genz1991},
denoted here as the corrected $\delta\gamma$ scheme for brevity, and a pairwise additive (PA) approximation of the HIs. The latter 
becomes exact at very low concentrations, but its prediction for $d_C^S$ worsens when protein volume fractions
$\phi \gtrsim 0.05$ are considered (see our discussion of \figurename~\ref{fig:S0_Dcoll_Exp_Theo} in Sec.~\ref{sec:Dynamic_Exp_vs_Theo}).
On the other hand, the PA predictions for $\eta_\infty$, and for the short-time self-diffusion coefficient $d_S$ not considered here,
are reliable up to substantially larger volume fractions, as has been ascertained in comparison to Stokesian Dynamics 
computer simulations \cite{Banchio2008, Heinen_TheoArticlePreparation} and experimental data \cite{heinen2010short}.
The PA expression for $d_C^S$ reads
\begin{eqnarray}\label{eq:Sedim_PA}
\dfrac{d_C^S}{d_0} &=& \dfrac{1}{S(q \to 0)} \left\lbrace \dfrac{d_S}{d_0} - 5\phi + 12\phi\int_1^\infty dx x \left[g(x)-1 \right] \right. \nonumber\\
&&+ 24 \phi\int_1^\infty dx x^2 g(x) \tilde{y}_{12}^a(x)\nonumber\\
&&+ \left.8 \phi \int_1^\infty dx x^2 g(x) \left[\tilde{x}_{12}^a(x)-\tilde{y}_{12}^a(x) \right] \right\rbrace, 
\end{eqnarray}
with $d_S$ given in PA approximation by 
\begin{equation}
\dfrac{d_S}{d_0} = 1 + 8 \phi \int_1^\infty dx x^2 g(x) \left[ x_{11}^a(x) + 2 y_{11}^a(x) -3\right].\label{eq:Ds_pa}\\
\end{equation}
The two-body mobility functions, $x_{ij}^a$ and $y_{ij}^a$, can be expanded analytically in powers of $\sigma/r = 1/x$. The short-range mobility parts 
\begin{eqnarray*}
\tilde{x}_{12}^a(x) &=& x_{12}^a(x) - (3/4) \, x^{-1} + (1/8)  \, x^{-3},\nonumber\\
\tilde{y}_{12}^a(x) &=& y_{12}^a(x) - (3/8) \, x^{-1} - (1/16) \, x^{-3},\nonumber
\end{eqnarray*}
include all terms in the series expansion in $1/x$ with the far-field terms up to the dipolar level subtracted off.
For $x > 3$, an explicit analytical expansion to $\mathcal{O}(x^{-20})$ is used  \cite{Schmitz1988}.
Since the series expansion in $1/x$ converges slowly at small separations, accurate numerical tables,
which account for lubrication at near-contact distances \cite{jeff_oni:84}, are employed for $x < 3$.    

The only input required in Eqs.~\eqref{eq:Sedim_PA} and \eqref{eq:Ds_pa} is the radial distribution function $g$, related to
$S$ by a one-dimensional Fourier transform \cite{Hansen_McDonald1986}. The two functions are obtained in our analysis by the analytical MPB-RMSA.

The second short-time method used in the present work for calculating $d_C^S \approx d_C$ and $\eta_\infty$, is the self-part
corrected $\delta\gamma$ scheme. In this scheme, $d_C^S$ is obtained from the exact relation \cite{Nagele1996} 
\begin{equation}\label{eq:Dcoll_dg}
\dfrac{d_C^S}{d_0} \lim_{q\to0}S(q)  = \dfrac{d_S}{d_0} + \lim_{q\to0} H_d(q) 
\end{equation}
containing the distinct part, $H_d(q)$, of the so-called hydrodynamic function $H(q)$. The $\delta\gamma$ scheme of Beenakker and Mazur
provides an easy-to-use integral expression for $H_d(q)$, including $S(q)$ as
the only required input. The explicit form of the $\delta\gamma$-scheme expression for $H_d(q)$ is given in
\cite{Genz1991, Banchio2008} and will be thus not repeated here.    

Extensive comparisons with Stokesian Dynamics simulations \cite{Banchio2008}, and experiments on charged colloids \cite{Gapinski2006, heinen2010short}, and for small $\phi$
also with PA calculations, have shown that the $\delta\gamma$ scheme predictions for $H_d(q)$ are quite good for 
all concentrations up to the freezing transition value, even though the $\delta\gamma$ scheme involves hydrodynamic approximations at any concentration.
In particular, it disregards lubrication effects. Lubrication, however, is inconsequential for charge-stabilized particles where near-contact
configurations are unlikely.

Different from $H_d(q)$, the accuracy of the $\delta\gamma$ scheme is less good for charged particles regarding the self-part, $d_S$, of $d_C^S$ in Eq.~\eqref{eq:Dcoll_dg}
\cite{Banchio2008,heinen2010short}. To remedy this deficiency, we use a hybrid method,
referred to as the self-part corrected $\delta\gamma$ scheme, in which $d_S$ is calculated using the PA expression in Eq.~\eqref{eq:Ds_pa}.
It has been shown both for charged colloids \cite{Banchio2008,heinen2010short,Heinen_TheoArticlePreparation} and Apoferritin
protein solutions \cite{Patkowski2005}, that this hybrid method works quite well at fluid state concentrations.

\subsection*{High-frequency viscosity}

The high-frequency viscosity, $\eta_\infty$, linearly relates the average suspension shear stress to the average
rate of strain in a low-amplitude, high-frequency oscillatory shear experiment. While this short-time quantity
has been rather routinely determined for micron-sized charge-stabilized colloids \cite{Bergenholtz1998_ExpGSE, Bergenholtz1998GSE}, a direct 
mechanical measurement of $\eta_\infty$ for BSA solutions is difficult, since the required frequencies
$\omega \gg \tau_I^{-1}$ are in the MHz regime. We are interested here in $\eta_\infty$ since, according to Eq.~\eqref{eq:eta_stat},
it is an important contribution to the static
viscosity $\eta$. The latter has been determined experimentally in the present work.  

In PA approximation, $\eta_\infty$ is given by \cite{Batchelor1972, Russel1984, Banchio2008}
\begin{equation}
\frac{\eta_\infty}{\eta_0} =1 + \frac{5}{2}\phi(1+\phi) + 60\phi^2\int_1^\infty dx x^2 g(x) J(x),\label{eq:eta_inf_PA}
\end{equation}
where the rapidly decaying shear mobility function $J(x)$, with $J(x) = 15/128~x^{-6} + \mathcal{O}(x^{-8})$ for stick boundary conditions, 
accounts for two-body HI effects. In performing the integral over $g(x)$, the leading-order long-distance contribution is dominating for $x>3$.
Accurate, numerical tables, where the lubrication effect for $x \approx 1$ is included, are used for $x<3$ \cite{jeff_oni:84}. 

The $\delta\gamma$ scheme of Beenakker and Mazur can be also used for calculating $\eta_\infty$. 
Similar to the $\delta\gamma$-scheme expression for $d_C^S$, the standard (2$^\text{nd}$ order) $\delta\gamma$ scheme result for $\eta_\infty$ consists of  
a microstructure-independent self-part, and a distinct part given in form of an integral over $S(q)$ \cite{Beenakker1984}.
In recent work, two of the present authors have shown that a self-part corrected
version of the original $\delta\gamma$ scheme expression for $\eta_\infty$ gives results for charged particles in very good agreement with Stokesian Dynamics
simulations \cite{Heinen_TheoArticlePreparation}. This self-part corrected $\delta\gamma$ scheme for $\eta_\infty$ is used
in the present work.

\subsection*{Static shear-viscosity}

In long-time rheological measurements on protein solutions under steady shear, there is an additional shear-stress relaxation part, $\Delta\eta$,
contributing to the static viscosity $\eta = \eta_\infty + \Delta\eta$. This contribution is influenced both by HIs and direct interaction forces.
It can be calculated approximately within the mode-coupling theory (MCT) of Brownian systems.
While a version of MCT for $\Delta\eta$ with far-field HI included has been discussed in earlier work together with an
extension to multicomponent systems \cite{Bergenholtz1998}, for analytical simplicity we use here the standard one-component
expression 
\begin{equation}\label{eq:Delta_eta_MCT_noHI}
\Delta\eta^{MCT} = \dfrac{k_B T}{60\pi^2}\int_0^\infty dt\int_0^\infty dq~q^4 {\left[\dfrac{S(q,t)}{S(q)}\dfrac{d}{dq}\log S(q)\right]}^2,
\end{equation}
which has been obtained, e.g. in  \cite{Bergenholtz1998}, under the neglect of HIs. In principle, $\Delta\eta^{MCT}$ should be calculated
self-consistently by a numerically expensive algorithm in combination with the corresponding MCT memory equation for $S(q,t)$ \cite{Banchio1999}.
However, the BSA solutions explored here are rather weakly coupled particle systems, with structure factor maxima $S(q_m) < 1.2$. Thus, as
we have thoroughly checked in comparison to fully self-consistent MCT calculations,
$\Delta\eta^{MCT}$ can be obtained more simply in a first iteration step where $S(q,t)$ in the integral of Eq.~\eqref{eq:Delta_eta_MCT_noHI}
is approximated by its short-time form $S(q,t)/S(q) = \exp[-q^2 d_0 t / S(q)]$, valid without HI. The difference to the fully self-consistent result for $\Delta\eta^{MCT}$
is at most a few percent, even for the most concentrated systems considered.

Moreover, again due to the only moderately strong interparticle correlations, $\Delta\eta$ augments $\eta_\infty$ by at most ten percent.
Therefore, the neglect of HI in $\Delta\eta^{MCT}$ can be expected to be rather insignificant for the systems considered since the dominant effect of HI is 
included already in $\eta_\infty$. Theoretical results for $\eta$ shown in this paper are all based on the first iteration solution
for $\Delta\eta^{MCT}$, and on $\eta_\infty$ calculated using the self-part corrected $\delta\gamma$ or PA schemes. For all
explored systems, the difference in $\eta_\infty$ between the PA and corrected $\delta\gamma$ scheme is at most two percent.    

\section{Static properties: experiment and theory}\label{sec:Static_Exp_vs_Theo}

\subsection{Form factor fit}
\begin{figure}
\centering
\includegraphics[width=.35\textwidth,angle=-90]{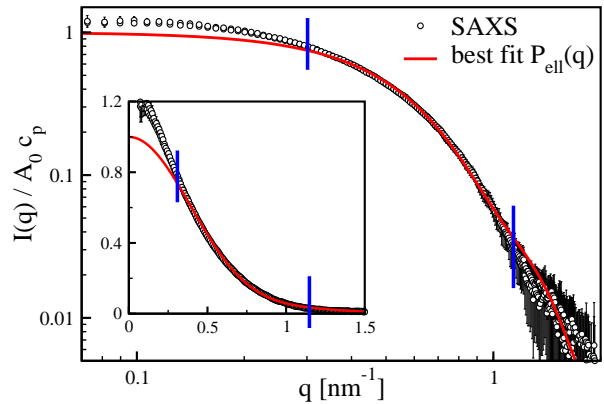}
\vspace{-1em}
\caption{BSA form factor fit. Open circles: SAXS intensities at two protein concentrations of $c_{p}=1$ and $2$ $g/l$, for $150$ mM of added NaCl. The SAXS intensities 
have been divided by $c_p$, and by a common, $q$-independent factor $A_0$. Red solid line: Angular-averaged spheroid form factor according to Eq.~\eqref{eq:Ellipse_Formfact},
fitted to the SAXS data within $0.3~\text{nm}^{-1} < q < 1.15~\text{nm}^{-1}$, as indicated 
by the blue vertical line segments. The obtained fit values are $a =1.75$ nm and $b =4.74$ nm. Inset: Intensity on a double linear scale.}
\label{fig:Formfakt_fit}
\end{figure}
In \figurename~\ref{fig:Formfakt_fit}, SAXS intensities for BSA solutions of very small protein weight concentrations, $c_p =1$ and $2$ mg/ml,
and $150$ mM of added NaCl, are shown along with the best-fit spheroid form factor. Note that our form factor fit relies on a simplified shape model,
so that some controlled systematic deviations from experimental data are to be expected.
To check for a residual effect of interparticle correlations on $I(q)$, $S(q)$ was calculated for the present two systems to first order in $\phi$
using the full DLVO potential, with $|Z| \sim 30$ and a Hamaker constant of $3 k_B T$ \cite{Lenhoff1996}.
The so-obtained structure factor deviates only very little from unity with $S(q\to0) \approx 0.99$. If vdW attraction
is ignored, $S(q\to0)$ is slightly lowered to $0.98$.

Thus, to fit the measured intensity in \figurename~\ref{fig:Formfakt_fit}, we have used Eq.~\eqref{eq:Iq_decoupling} for $I(q)$ with $S_m(q)$ set equal to one.
Using an automatic weighted least-squares minimizer, the spheroid semi-axes $a$ and $b$ entering into $P_{ell}(q)$ were varied to achieve a best
fit intensity for a given prefactor $A$ in Eq.~\eqref{eq:Iq_decoupling}. 
This fitting procedure was iterated for different values for $A$, until optimal agreement with the SAXS intensities within
the range $0.3~\text{nm}^{-1} < q < 1.15~\text{nm}^{-1}$  was achieved, resulting in $a =1.75$ nm and $b =4.74$ nm.
These values for the spheroid semi-axes are in good accord with previously reported values, 
and in reasonable agreement with the linear dimensions of the reported heart-shape like crystal structure of albumins \cite{Carter1994, Ferrer2001, Leggio2008, Zhang2007}.
In a related, recent study by part of the present authors \cite{Roosen-Runge2011}, similar values
$a = 1.80 \pm 0.05$ nm and $b = 4.60 \pm 0.15$ nm have been determined, which are in decent agreement with the values obtained here. 
The optimized value for $A$, denoted by $A_0$, has been also used in our SAXS intensity fits for systems without added salt,
which will be discussed in the following subsection.

The best-fit form factor, $P_{ell}$, depicted in \figurename~\ref{fig:Formfakt_fit} deviates from the SAXS intensities outside the fitted $q$-range.
For $q \gtrsim 1.15~\text{nm}^{-1}$, corresponding to length scales $2\pi/q \lesssim 6$ nm $\lesssim \sigma$, 
the complex internal structure of BSA is probed, which is not accounted for in our simplifying SY model. The deviations visible for $q \lesssim 0.3~\text{nm}^{-1}$, 
corresponding to distances of roughly $20$ nm or larger, are likely due to additional scattering species made up of larger particles such as BSA oligomers
or impurities.
Since the size-, form-, and charge-distributions of oligomers and impurities are unknown, our choice of the lower $q$-boundary in fitting $I(q)$ is somewhat more
ambiguous than the upper boundary. Therefore, we have repeated the intensity fitting for various low-$q$ boundaries, finding that the weighted
least squares deviation increases dramatically if the boundary is selected below $0.3~\text{nm}^{-1}$. Moreover, the fit values for $a$ and $b$ remain essentially
constant when the lower $q$-boundary is chosen larger than $0.3~\text{nm}^{-1}$.      

The fit parameters of a spheroid form factor to SAXS data of proteins in general depend slightly on the measured $q$ range,
the prepared protein concentration, solvent and salt conditions, and background subtraction.
In the context of the present study, the related changes of the spheroid model parameters are small compared to the experimental
error bars and will be discussed in the next section.

\subsection{Concentration series of scattered intensities}
\begin{figure}[h]
\centering
\includegraphics[width=.4\textwidth]{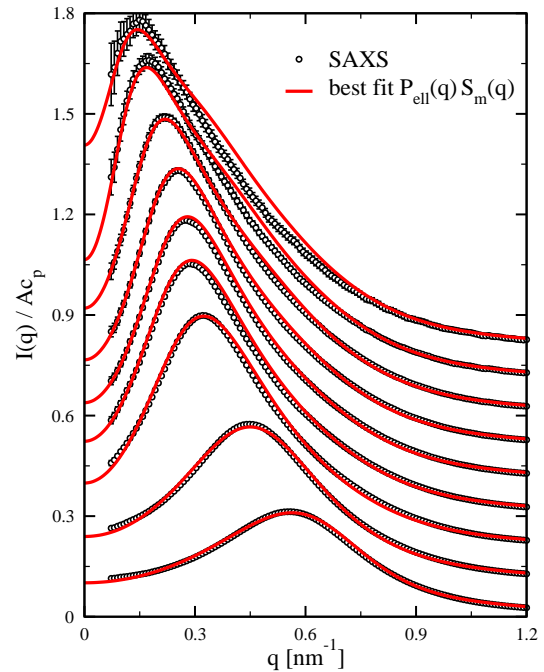}
\vspace{-1em}
\caption{SAXS intensities from BSA solutions at various $c_p$, without added salt, divided by $A c_p$. From top to bottom: $c_p = 0.9, 1.8, 4.5, 7.2, 9, 13.5, 18, 45,$ and $90$ mg/ml.
The intensity curves are displaced in steps of $0.1$ along the vertical axis for better visibility.
The SAXS-data for the extended range $q \lesssim 4$ $\text{nm}^{-1}$ were taken into account in all fits, but shown here only up to $q = 1.2$ $\text{nm}^{-1}$.
Red solid lines: best fits according to Eq.~\eqref{eq:Iq_decoupling} with $S(q)$ calculated in MPB-RMSA.
The fit parameters are listed in \tablename~\ref{tab:lowsalt_fitparams}.}
\label{fig:SAXS_Intens_fit}
\end{figure}
\begin{table}[h]
\small
  \caption{Fit parameters $\phi$, $|Z|$, and $A/A_0$, for the BSA concentration series without added NaCl, with intensities shown in \figurename~\ref{fig:SAXS_Intens_fit}.
The additional parameters $L_B =0.711$ nm, $\sigma =\sigma_{B_2} =7.40$ nm, $a =1.75$ nm, $b =4.74$ nm are kept fixed,
and $A_0$ is taken from the form factor fit in \figurename~\ref{fig:Formfakt_fit}. The obtained 
fit values at $c_p =0.9, 1.8, 45,$ and $90$ mg/ml should be taken with a pinch of salt (see related text). The listed values for $c_p$ are 
according to Eq.~\eqref{eq:cp_exp}.}
  \label{tab:lowsalt_fitparams}
  \begin{tabular*}{0.485\textwidth}{@{\extracolsep{\fill}}lllll}
    \hline
    \textbf{~$\mathbf{c_p}$ [mg/ml]~} & \rule[-3mm]{0mm}{8mm}~$\boldsymbol{\phi}$~ & ~$\mathbf{|Z|}$~ & ~\textbf{$\mathbf{n_s}$ [$\mathbf{\mu}$M]}~ & ~$\mathbf{A/A_0}$~\\
    \hline
~$0.9$~   & ~$5.19\times10^{-4}$~\rule[-2mm]{0mm}{6mm} & ~$34.5$~ & ~1216~ & ~1.20~\\ 
~$1.8$~   & ~$1.34\times10^{-3}$~\rule[-2mm]{0mm}{6mm} & ~$18.8$~ & ~608~  & ~1.08~\\ 
~$4.5$~   & ~$3.72\times10^{-3}$~\rule[-2mm]{0mm}{6mm} & ~$19.1$~ & ~1278~ & ~0.96~\\ 
~$7.2$~   & ~$6.97\times10^{-3}$~\rule[-2mm]{0mm}{6mm} & ~$16.7$~ & ~1497~ & ~0.97~\\ 
~$9$~  & ~$1.04\times10^{-2}$~\rule[-2mm]{0mm}{6mm} & ~$14.6$~ & ~1510~ & ~1.05~\\    
~$13.5$~  & ~$1.28\times10^{-2}$~\rule[-2mm]{0mm}{6mm} & ~$12.6$~ & ~1297~ & ~0.81~\\ 
~$18$~  & ~$2.06\times10^{-2}$~\rule[-2mm]{0mm}{6mm} & ~$10.8$~ & ~1292~ & ~0.85~\\   
~$45$~  & ~$8.19\times10^{-2}$~\rule[-2mm]{0mm}{6mm} & ~$9.47$~  & ~2375~ & ~1.0~\\   
~$90$~ & ~$1.74\times10^{-1}$~\rule[-2mm]{0mm}{6mm} & ~$8.52$~  & ~3323~ & ~1.0~\\    
    \hline
  \end{tabular*}
\end{table}

\figurename~\ref{fig:SAXS_Intens_fit} includes the SAXS intensities for all explored BSA solutions without added salt that could be fitted using 
the decoupling approximation expression in Eq.~\eqref{eq:Iq_decoupling}, for $S(q)$ calculated in MPB-RMSA using the screened Coulomb potential
in Eq.~\eqref{eq:HSY_pair_pot}.
In order to emphasize the shape differences across the dataset, the intensities are divided by their respective fitted amplitudes $A$,
and by the protein concentrations $c_p$.
The most concentrated solution shown here is the one for $c_p =90$ mg/ml.
Two even more concentrated systems for $c_p =180$ and $270$ mg/ml are not depicted in the figure since
their intensities could not be fitted reasonably well by the SY model.

In order to fit the experimental intensity data using Eq.~\eqref{eq:Iq_decoupling}, some deviations of the prefactor $A$ from
the optimized form factor fit value $A_0$ have to be allowed for (see \tablename~\ref{tab:lowsalt_fitparams}).
The fit of each individual intensity curve in \figurename~\ref{fig:SAXS_Intens_fit} was made as follows:
After dividing the SAXS intensity by $A_0$ and $c_p$, the 
weighted sum of quadratic deviations between SAXS data points and the intensity according to Eq.~\eqref{eq:Iq_decoupling} was minimized by an automatic
three-dimensional weighted least-squares minimizer
with respect to the fitting parameters $\left\lbrace|Z|, n_s, \phi\right\rbrace$. For each concentration, the whole 
experimental dataset was used, for wavenumbers from $0.07$ to about $4$ $\text{nm}^{-1}$. If the fit was unsatisfactory, the
prefactor $A$ was slightly altered, and the optimization with respect to $\left\lbrace|Z|, n_s, \phi\right\rbrace$ was repeated. This procedure was iterated
until convergence in all fit parameters was achieved. For all considered concentrations, $L_B =0.711$ nm, 
$\sigma =\sigma_{B_2} =7.40$ nm, $a =1.75$ nm, and $b =4.74$ nm  were kept fixed.
\tablename~\ref{tab:lowsalt_fitparams} summarizes the obtained best fit parameters.

While the overall intensity fits for the two lowest concentrations, $c_p =0.9$ and $1.8$ mg/ml, look quite reasonably good, they contain some peculiarities.
A shoulder is present in the fit intensity extending from $q \approx0.3$ to $0.8$ $\text{nm}^{-1}$,
overshooting the experimental data by several standard deviations. Moreover, the prefactor $A$ is substantially larger 
than $A_0$ in both cases, and the fitted effective charge number $|Z|$ assumes a questionably large value of $34.5$ for $c_p =0.9$ mg/ml.
These peculiarities can be attributed to impurity contributions neglected in Eq.~\eqref{eq:Iq_decoupling}.
Note also that the maximal intensities in both systems occur at wavenumbers well below the value $0.3$ $\text{nm}^{-1}$,
where impurities are found to obstruct also the form factor fit in \figurename~\ref{fig:Formfakt_fit}.

All our attempts to remedy these fitting problems for the two most dilute samples failed. Lacking information about the shape and size distribution,
and the interactions of the impurities, we cannot improve on Eq.~\eqref{eq:Iq_decoupling}.
Restricting the wavenumber interval in the fitting procedure to $q \gtrsim0.3$ $\text{nm}^{-1}$ leads to no improvement, either.
While Eq.~\eqref{eq:Iq_decoupling} is expected to be quite accurate in this restricted $q$-range, the maximum in $I(q)$ is not included.
The intensity for $q >0.3$ $\text{nm}^{-1}$ is a
monotonically decaying curve, almost completely determined by the form factor. It therefore lacks distinct features coming from
particle correlations, rendering the fit
with respect to $\left\lbrace|Z|, n_s, \phi\right\rbrace$ into an overdetermined problem.       
For all these reasons, our fit parameters in \tablename~\ref{tab:lowsalt_fitparams} for $c_p =0.9$ and $1.8$ mg/ml should not be considered as quantitatively accurate.
 
Except for the two most dilute systems, all other systems with concentrations from $c_p =4.5$ to $90$ mg/ml included in \figurename~\ref{fig:SAXS_Intens_fit}
can be excellently fitted by Eq.~\eqref{eq:Iq_decoupling}. The obtained effective charges, salt concentrations, and 
volume fractions all assume reasonable values, showing systematic dependencies on the BSA concentration. Note, however, that for
$c_p =45$ and $90$ mg/ml, the SY model is pushed to its limit. On assuming a Hamaker constant of $3$ $k_B T$ \cite{Lenhoff1996},
the repulsive barrier height of the DLVO potential becomes very small, with values of $1.3$ and $0.5$ $k_BT$ at $c_p =45$ and $90$ mg/ml, respectively.
The contact value of $g(x)$ at $x=1$ is just barely zero for the more dilute system, whereas $g(x=1^+) \approx 0.9$ in the more concentrated system.
Obviously, the SY model with purely repulsive, spherically symmetric pair interactions is bound to fail when the particles are allowed to come
into hard-core contact. Thus, the system with $c_p =45$ mg/ml, and fitted volume fraction $\phi =8.19\%$,
is clearly on the borderline of the SY model. Somewhat unexpectedly, and probably fortuitously, the system with $c_p =90$ mg/ml can still
be fitted with good accuracy. Summarizing, the fit values for the most concentrated systems with $c_p =45$ and $90$ mg/ml in \tablename~\ref{tab:lowsalt_fitparams}
should be interpreted with caution, since the fit parameters might be significantly distorted by the discussed deficiencies of the SY model.
An indication for this could be the obtained fit values for $\phi(c_p)$, which for
the two most concentrated samples clearly overshoot the linear dependence on $c_p$ found approximately for the lesser concentrated
systems (see \tablename~\ref{tab:lowsalt_fitparams}).    

In closing our discussion of the static scattered intensities, we note that fit parameters slightly different from the ones in \tablename~\ref{tab:lowsalt_fitparams}
are obtained, when in place of the BSA model spheroid axes $(a,b) = (1.75~\text{nm}, 4.74~\text{nm})$, the values $(a,b) = (1.80~\text{nm}, 4.60~\text{nm})$
given in \cite{Roosen-Runge2011} are used. For instance, at $c_p = 4.5$ and $18$ mg/ml, the best-fit values for $|Z|$ change to $18.4$ and $10.7$, respectively.
Note that, in comparison to \cite{Zhang2007}, where the RMSA was employed in fitting $I(q)$, we use here the improved MPB-RMSA integral equation scheme for $S(q)$,
resulting in more precise fit-values. Moreover, different from the earlier intensity fitting described in \cite{Zhang2007}, the dephasing influence on $I(q)$
originating from the particle asphericity is accounted for approximately in the decoupling approximation used in the present study.
The slightly different spheroid semi-axes  $(a,b) = (1.80~\text{nm}, 4.60~\text{nm})$, and the corresponding,
slightly changed fit-parameters, do not cause appreciable changes in the dynamical properties. For instance, the
collective diffusion coefficient changes by no more than $3\%$, and the changes in the static- and high-frequency viscosities are less
than $0.1\%$. 
Note that the somewhat smaller spheroid causes changes of the fitted volume fraction of about 5\% which does not change absolute values but slightly
rescales the protein concentration axis for the theoretical predictions.

\section{Dynamic properties: experiment and theory}\label{sec:Dynamic_Exp_vs_Theo}

In the following, we compare the DLS data for the collective diffusion coefficient of BSA solutions, and the static shear viscosity measured in our
suspended couette-type rheometer,
to the results of the dynamic schemes discussed in Sec. \ref{sec:Theory}.
Moreover, we test the validity of a generalized Stokes-Einstein relation connecting the viscosity to the collective diffusion coefficient and the
isothermal osmotic compressibility.
We reemphasize here that the employed theoretical schemes use
$S(q)$ and $g(r)$ as the only input. With $S(q)$ and $g(r)$ determined from the fits to the SAXS-intensities, all theoretical results
for $d_C$, $\eta_\infty$ and $\eta$ are thus obtained without any additional adjustable parameters.

\subsection{Collective diffusion coefficient}

\begin{figure}
\centering
\includegraphics[width=.4\textwidth]{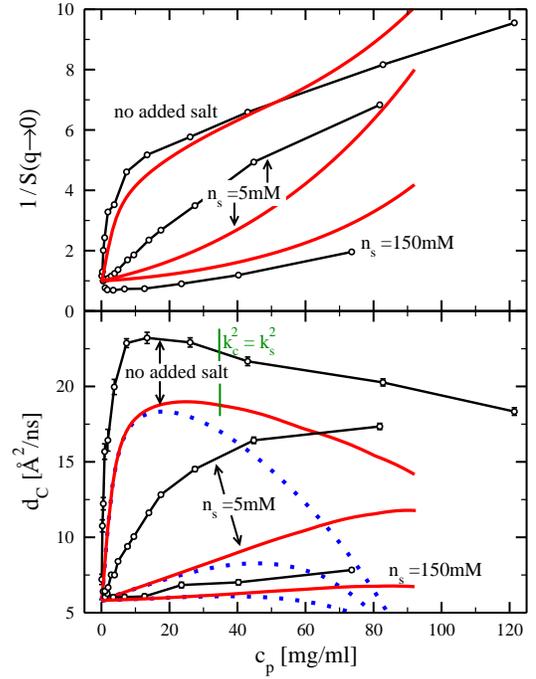}
\vspace{-1.5em}
\caption{Top panel: Inverse zero-wavenumber limiting static structure factor of BSA solutions, obtained from SLS (connected black circles) 
and our MPB-RMSA scheme (red solid lines). Number concentrations, $n_s$, of added NaCl as indicated. 
Bottom panel: Fast mode coefficient, $D_1 = d_C^L$, obtained from the discussed double-exponential fit to the DLS data of BSA solutions
(connected black circles), and $d_C^S$ calculated by the 
self-part corrected $\delta\gamma$ scheme (red solid lines), and the PA approximation (blue dotted curves).
All theoretical curves are based on input parameters $\phi(c_p)$ and $Z(c_p)$ interpolated from \tablename~\ref{tab:lowsalt_fitparams}. In the zero added-salt case,
the $n_s(c_p)$ values were also interpolated using \tablename~\ref{tab:lowsalt_fitparams}. Theoretical results for added NaCl are obtained using fixed salt
concentrations of $n_s =5$ and $150$ mM. The input parameters $L_B =0.711$ nm, 
$\sigma =\sigma_{B_2} =7.40$ nm, $a =1.75$ nm, $b =4.74$ nm, and $d_0 =d_0^{ell}(a,b) =5.82~\text{\AA}^2/\text{ns}$ are kept fixed throughout.
For the zero added-salt case, the green vertical line segment at $c_p \approx 34$ mg/ml marks the protein mass-concentration where 
the surface-released counterion contribution to $k^2$ in Eq.~\eqref{eq:DLVOSscreeningConst} is equal to the coion contribution.}
\label{fig:S0_Dcoll_Exp_Theo}
\end{figure}

\figurename~\ref{fig:S0_Dcoll_Exp_Theo} includes our SLS/DLS data for $1/S(q\to0)$ (upper part) and $d_C^L = d_C$ (lower part),
for aqueous BSA solutions in comparison with the theoretical predictions. Systems without added salt, and for concentrations $n_s =5$ and $150$ mM of added NaCl,
are considered. Additional measurements using $500$ mM of added NaCl (data not shown) agree almost perfectly with the data
for $n_s =150$ mM, indicating that electrostatic repulsion is fully screened already at $n_s =150$ mM.   
As the input to the dynamics schemes, $S(q)$ and $g(r)$ were generated by the MPB-RMSA, using concentration-interpolated input parameters $\phi(c_p)$ and $Z(c_p)$
based on \tablename~\ref{tab:lowsalt_fitparams}. For no added salt, $n_s(c_p)$ was interpolated using \tablename~\ref{tab:lowsalt_fitparams},
while $n_s =5$ and $150$ mM were kept
fixed (independent of $c_p$) in the corresponding theoretical calculations. The value $d_0^{ell} =5.82$ ${\text{\AA}}^2/\text{ns}$
of the spheroid translational free diffusion coefficient was used to obtain $d_C$ in the experimental units from the dimensionless results for $d_C/d_0$
obtained by both theoretical schemes.

For no added salt, the experimental $d_C(c_p)$ assumes a maximum at $c_p \approx 10 - 20$ mg/ml. This maximum is qualitatively
reproduced by both theoretical schemes (corrected $\delta\gamma$ and PA), but its location is predicted to occur at somewhat larger concentrations 
$c_p \approx 20 - 30$ mg/ml. For BSA concentrations larger than the concentration at the maximum value for $d_C$, the PA-predicted $d_C(c_p)$
reduces strongly, eventually reaching unphysical negative values for $c_p \gtrsim 110$ mg/ml. This illustrates the expected failure of the PA scheme at higher concentrations,
indicating that three-body contributions to HI, totally left out in the PA, but not in the $\delta\gamma$ scheme, come into play for $c_p \gtrsim 30$ mg/ml.
Up to the concentration value at the maximum of $d_C$, both schemes agree very well, with residual differences
not visible for $c_p \lesssim 20$ mg/ml on the scale of \figurename~\ref{fig:S0_Dcoll_Exp_Theo}.
Despite its residual small inaccuracies, the self-part corrected $\delta\gamma$ expansion will therefore be used in the following calculations of $d_C$.

The physical origin of the non-monotonous concentration dependence $d_C(c_p)$ at low concentrations 
of salt can be understood on the basis of Eq.~\eqref{eq:Sedim_PA}, rewritten using $d_C \approx d_C^S$ as
\begin{equation}\label{eq:Dcoll_from_H0_S0}
\dfrac{d_C}{d_0} = \lim_{q\to0} \dfrac{H(q)}{S(q)}, 
\end{equation}
with $H(q) = d_s/d_0 + H_d(q)$. The ratio in Eq.~\eqref{eq:Dcoll_from_H0_S0} 
consists of two competing factors. The factor \mbox{$1/S(q \to 0)$}, inversely proportional to the isothermal osmotic compressibility of ideally monodisperse
particles, increases monotonically as a function of the BSA concentration. Owing to the larger coupling constant $\gamma$ in Eq.~\eqref{eq:DLVOcouplingConst},
a much steeper initial increase of $1/S(q\to0)$ is observed for weakly screened systems than for systems with added salt
(c.f. the top panel of \figurename~\ref{fig:S0_Dcoll_Exp_Theo}).
As $c_p$ is further increased, the amount of surface-released counterions increases correspondingly, leading to an enhanced electrostatic screening. As a consequence, the
rate of change of $1/S(q \to 0)$ with $c_p$ reduces significantly at a colloid concentration roughly set by the criterion, $k_c^2(c_p) = k_s^2$, of equal 
surface released counterion and salt-co-ion contributions to the screening parameter in Eq.~\eqref{eq:DLVOSscreeningConst}.

The nominator in Eq.~\eqref{eq:Dcoll_from_H0_S0} is the reduced
sedimentation velocity, $H(q \to 0)$, which is known from theory and experiment \cite{heinen2010short}
to decrease monotonically, for not too large concentrations and low salinity according to $1 - a_{sed}~\phi^{1/3}$, with $a_{sed} =1.6~-~1.8$
in the case of highly charged particles, and as
$1 - 6.546~\phi + 21.918~\phi^2 + \mathcal{O}(\phi^3)$ for neutral hard spheres \cite{Cichocki2002}.
For strongly correlated particles, the competition between decreasing compressibility
and decreasing sedimentation coefficient with increasing $c_p$ leads thus to a maximum in $d_C(c_p)$, at
a concentration roughly determined from $k_c^2(c_p) = k_s^2$.

The DLS-measured values for $d_C$ are not quantitatively reproduced by the self-part corrected $\delta\gamma$ scheme.
Both in the zero added-salt case, and for $n_s =150$ mM,
$d_C$ is underestimated by the corrected $\delta\gamma$ scheme prediction by about $25\%$.
The difference might be simply due to the complex-shaped BSA proteins having
a translational free diffusion coefficient larger than the value $d_0^{ell} =5.82$ ${\text{\AA}}^2/\text{ns}$ used in the SY model.
In fact, an extrapolation of the experimental data for $d_C$
to zero concentration leads to a larger value for $d_0$ in the range of $6 - 7$ ${\text{\AA}}^2/\text{ns}$, which can completely explain the differences in $d_C$
between experiment and theory. However, this low-concentration extrapolation should not be over-interpreted as being conclusive, since
the experimental data are rather noisy for low concentrations.

While the agreement between the theoretical and the experimental $d_C$'s is overall rather
satisfying for very low and very high salt content, strong differences are found for the intermediate added NaCl concentration of $5$ mM.
This is not surprising, however, since already the zero added-salt experiments led to fit values for $n_s$ of $1$ to $3$ mM.
Therefore, $n_s$ is most probably a function of $c_p$ also in the
$5$ mM added NaCl case, instead of being constant as assumed in the calculations.
Moreover, there is no obvious reason to expect that the relation $Z(c_p)$, interpolated from \tablename~\ref{tab:lowsalt_fitparams},
remains valid at arbitrary added salt concentrations.
Additional future SAXS measurements at $5$ mM added NaCl are
necessary to determine, for this case, the precise dependence of $n_s$ and $Z$ on $c_p$.

\subsection{Static viscosity}

\begin{figure}
\begin{center}
\vspace{-1em}
\includegraphics[width=.35\textwidth,angle=-90]{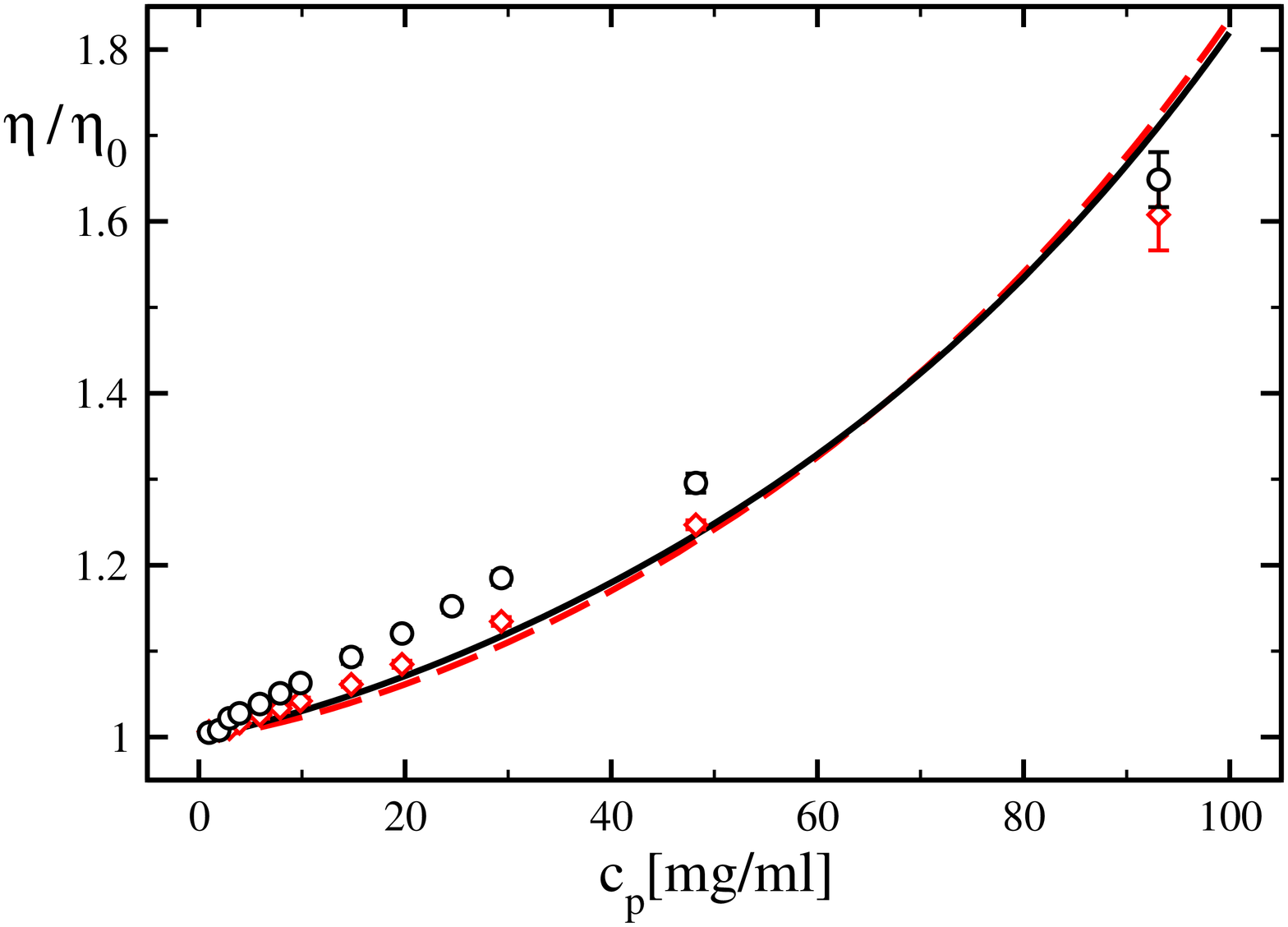}\\
\vspace{-1.5em}
\includegraphics[width=.35\textwidth,angle=-90]{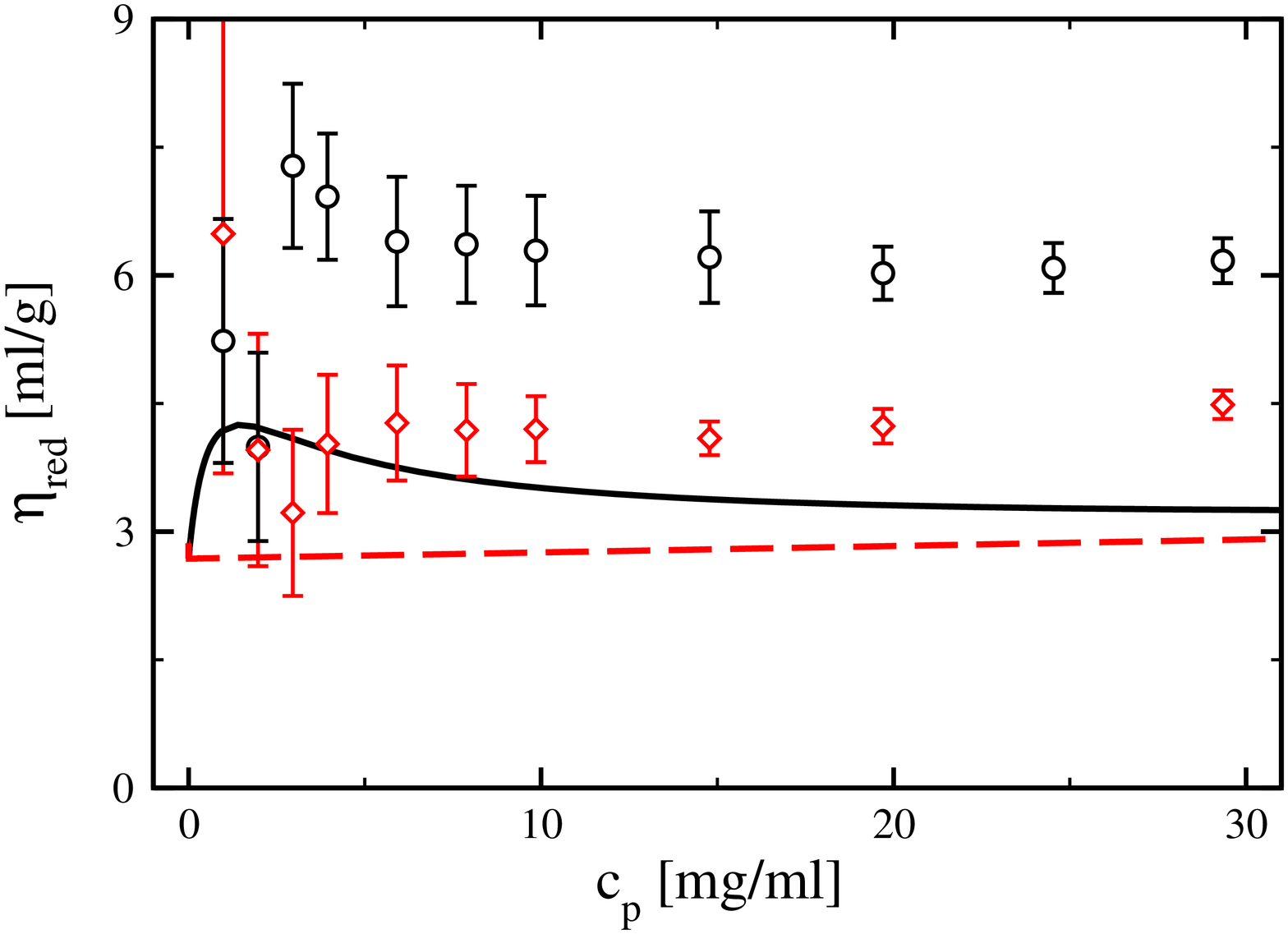}
\vspace{-2em}
\end{center}
\caption{Static relative (top panel) and reduced (bottom panel) viscosity for $T =25^\circ$ C as function of $c_p$.
Theoretical curves are based on input parameters $\phi(c_p)$ and $Z(c_p)$, concentration-interpolated using \tablename~\ref{tab:lowsalt_fitparams}.
Symbols: experimental data without added salt (black circles) and with $n_s =150$ mM (red diamonds).
Lines: theoretical results without added salt (black solid line, $n_s(c_p)$ interpolated using \tablename~\ref{tab:lowsalt_fitparams})
and with a fixed salt concentration of $n_s =150$ mM (red dashed curve). Note the different $c_p$-ranges in the two panels of the figure.}
\label{fig:Visc_vs_conc_Exp_Theo}
\end{figure}

The rheometric results for $\eta$ without added salt, and with $150$ mM of added NaCl, are
plotted in \figurename~\ref{fig:Visc_vs_conc_Exp_Theo} as a function of $c_p$, and compared to the theoretical predictions.
Apart from pronounced differences at lower concentrations, discussed in detail further down, the experimental data
agree overall decently well with the theoretical predictions. Due to the rather weak microstructural ordering of the BSA proteins,
characterized by structure factor peak heights less than $1.2$
even for the most concentrated samples, the shear-stress relaxation term $\Delta\eta$ contributes only little
to $\eta$, with a maximum relative contribution of about $10\%$ near $c_p =100$ mg/ml.
The dominant contribution to $\eta$ is given by $\eta_\infty$,
which is predicted to good accuracy both by the PA scheme and the corrected $\delta\gamma$ scheme, with practically equal results. The PA scheme
is applicable to the whole experimentally probed concentration range of $c_p \lesssim 100$ mg/ml, since three-body and
higher order HI contributions affect $\eta_{\infty}$ to a lesser extent than $d_C$ (c.f. here Fig. \ref{fig:S0_Dcoll_Exp_Theo}, showing the failure of the
PA prediction for $d_C$ already for $c_p \lesssim 50$ $g/l$).

The addition of larger amounts of salt lowers the values for $\eta$, as can be noticed from the two experimental data sets depicted
in \figurename~\ref{fig:Visc_vs_conc_Exp_Theo}.
The reason for this is the enhanced electrostatic screening, causing $\Delta\eta$ to decrease
with increasing salinity in going from strongly structured, charged spheres to basically neutral hard spheres.
In contrast, $\eta_\infty$ is known from theory and experiment \cite{Banchio2008} to \textit{increase}
upon the addition of salt, due to the enlarged influence of near-field HIs when the particles are allowed to get closer to each other in electrostatically
screened systems. Thus, $\eta_\infty$ and $\Delta\eta$ have opposite trends in their dependencies on the concentration of added salt. These 
competing trends are the reason for the weak crossover in the two theoretical curves for $\eta$,
noticed in the top panel of \figurename~\ref{fig:Visc_vs_conc_Exp_Theo} at $c_p \approx 67$ mg/ml.
For particle concentrations larger than this $c_p$ value,
the increase of $\eta_\infty$ overcompensates the decrease in $\Delta\eta$ when, in place of the zero added-salt system, a system with
$n_s =150$ mM is considered. That such a weak crossover is not observed in the
experiment data in \figurename~\ref{fig:Visc_vs_conc_Exp_Theo}, points to an underestimation of the crossover concentration by
our simplifying theories for $\eta$, possibly due to the neglect of HIs in the $\Delta\eta$ calculation.

A remarkable feature is noticed from the bottom panel of \figurename~\ref{fig:Visc_vs_conc_Exp_Theo}, where we plot the so-called reduced viscosity,
\begin{equation}\label{eq:red_visc}
\eta_{red}(c_p) = \dfrac{\eta(c_p) - \eta_0}{\eta_0 c_p}, 
\end{equation}
as a function of $c_p$. The function $c_p \eta_{red} / \phi$ reduces to the intrinsic viscosity, $[\eta]$ at very low volume fractions
where $\eta \to \eta_0 + [\eta]\phi$. Features of dilute systems are more clearly revealed in $\eta_{red}$ than in $\eta$.

Both experimental data sets in the bottom panel of \figurename~\ref{fig:Visc_vs_conc_Exp_Theo} show a local maximum of $\eta_{red}$ at low $c_p$ values,
which for the zero added-salt system (black open circles) is visible as a weak non-monotonicity near $c_p \approx 3$ mg/ml. For the system with $150$ mM added NaCl
(red open diamonds), the experimental maximum is represented essentially by 
a single data point at $c_p =1$ mg/ml, where $\eta_{red} \approx 6.5$ ml/g, whereas the remaining data points describe a nearly constant plateau value of $4.5$ ml/g.
This plateau value is in good overall agreement with reported values for $\eta_{red}$ at low $c_p$, in the range of $3.8$ to $4.9$ ml/g
\cite{Tanford1956, Kupke01081972, Placidi1998, curvale2008intrinsic}. 

Regarding the large experiment error bars at very low $c_p$, from the figure, we can not attribute physical significance to the single-point maximum in
the $n_s =150$ mM system. A more refined data resolution in a future experimental study is clearly needed here.
Even the maximum in $\eta_{red}$ for the zero added-salt case might be disputable on basis of the experimental
data alone. However, the existence of such a maximum in $\eta_{red}$ draws its credibility from the comparison
to the theoretical results, showing a maximum in $\eta_{red}(c_p)$ at a slightly lower value of $c_p$. A similar non-monotonic 
behavior of $\eta_{red}(c_p)$, with a pronounced peak at low $c_p$, has been measured also
in polyelectrolyte systems \cite{Forster1995,Forster1996,Eisenberg1977},
in low-salinity suspensions of charged silica spheres \cite{Okubo1987}, and in microgels \cite{Philipse2002}.
The effect has been described theoretically by scaling arguments \cite{Rabin1987}, by the Rice-Kirkwood equation \cite{Kirkwood1959} for the shear viscosity
in combination with a screened Coulomb potential \cite{Nishida2004}, and for rod-like particles using a MCT scheme similar to ours \cite{Yethiraj2004}.
In these earlier treatments, HI has been disregarded altogether. In our approach, HI is included in the for the present systems dominating
$\eta_\infty$ part of $\eta$.

To rule out that the non-monotonicity of the theoretical $\eta_{red}(c_p)$ is caused by BSA-specific dependencies
of $|Z|$ and $n_s$ on $c_p$, (c.f. \tablename~\ref{tab:lowsalt_fitparams}), we have investigated additionally a model system
for fixed $|Z| =20$ and $n_s =1$ mM, where we find again a maximum in $\eta_{red}(c_p)$. Thus, the maximum in $\eta_{red}(c_p)$ is 
a generic effect in weakly screened HSY fluids. It is entirely
due to the shear-stress relaxation term $\Delta\eta$, for $(\eta_\infty-\eta_0)/(\eta_0 \phi)$ increases monotonically in $c_p$
at arbitrary salt concentration.
Since the HIs are neglected in our MCT treatment of the shear-stress relaxation part $\Delta\eta$, we conclude that the local maximum in $\eta_{red}$
is basically a non-hydrodynamic effect, arising from electrostatic repulsion. We point out that the discussed physical mechanism underlying the non-monotonic
behavior of $\eta_{red}(c_p)$ is different from the one causing the maximum in $d_C$ as a function of $c_p$.
The latter maximum originates from a competition between electrostatic repulsion
and hydrodynamic slowing in crowded systems. It is therefore not surprising that the maxima 
in $\eta_{red}$ and $d_C$ are located at considerably different protein concentrations.
Whereas the maximum of $d_C$ occurs at $c_p \approx30$ mg/ml (c.f. \figurename~\ref{fig:S0_Dcoll_Exp_Theo}),
the maximum in $\eta_{red}$ is observed at $c_p \lesssim 5$ mg/ml.

The theoretical values for $\eta_{red}$ in \figurename~\ref{fig:Visc_vs_conc_Exp_Theo} underestimate the experimental data by a
factor of about $1/2$. In the low-concentration regime, the theoretical result for $\eta_{red} c_p / \phi$ approaches $[\eta] =2.5$,
owing to the underlying effective sphere model.
The intrinsic viscosity of BSA modeled as a spheroid is $[\eta]^{ell} =3.25$, which is larger than the
value for a sphere by a factor of $1.3$ only. Therefore, this can not be the only cause for the observed deviation.
However, the actual intrinsic viscosity of a heart-like shaped BSA protein is neither equal to that of a spheroid nor to that of an effective sphere. We recall
here our discussion of \figurename~\ref{fig:S0_Dcoll_Exp_Theo}, where we argued that $d_0$ for a BSA protein might well be about $25\%$ larger than
the free diffusion coefficient, $d_0^{ell}$, of the model spheroid. We can similarly argue that the 
observed differences between the experimental and theoretical $\eta_{red}$ may be largely due to a value for the intrinsic viscosity of BSA of about $4 - 5$,
which is $20 - 50\%$ larger than ${[\eta]}^{ell}$, and about twice as large as the $[\eta]$ value of spheres. This could explain the observed difference.

Finally, we note here that electrokinetic contributions to $\eta$, $d_S$ and $d_C$, originating from the non-instantaneous response of the 
microion-clouds around each protein, are not included in our treatment. Microion electrokinetics has the effect of lowering somewhat the values of $d_S$ and $d_C$
\cite{Retailleau1999,Patkowski2005}, while enlarging the viscosity $\eta$ \cite{Ohshima2007, Sherwood2007}.
These effects can be expected to be stronger when $\kappa^{-1}$ is approximately equal to the 
particle size. Furthermore, electrokinetic effects are expected to be less significant at higher protein concentrations \cite{McPhie2007, McPhie2008}.   

\subsection{Relation between viscosity and collective diffusion}

Kholodenko and Douglas \cite{Kholodenko1995} have proposed the approximate generalized Stokes-Einstein (GSE) relation
\begin{equation}\label{eq:KD-GSE}
\dfrac{d_C(\phi)\eta(\phi)}{d_0 \eta_0}\sqrt{S(q\to0,\phi)} \approx 1,
\end{equation}
between collective diffusion coefficient, (static) viscosity and the square-root of the isothermal osmotic compressibility coefficient $S(q\to0,\phi)$.
If this relation were exactly valid, the dimensionless function on the left-hand-side (lhs) of Eq.~\eqref{eq:KD-GSE} would be a constant equal to one.
The (approximate) validity of a GSE relation would be very useful from an experimental viewpoint, since it allows to infer
viscoelastic properties such as $\eta_\infty$ and $\eta$ from a dynamic scattering experiment where diffusion 
coefficients are determined. This is of particular relevance when the amount of protein available is too small for a mechanical rheometer measurement.
Since we have experimental data sets for $\eta$, $d_C$, and $S(q\to0)$ for BSA solutions with low and high salt content at our disposal,
together with theoretical tools to calculate these properties, we are in the position to scrutinize the accuracy of the KD-GSE
relation. We can do this not only for the special case of BSA solutions, but with our theoretical methods more generally for
arbitrary spherical colloidal particles interacting by the HSY potential in Eq.~\eqref{eq:HSY_pair_pot}.   

In their discussion of the GSE relation in Eq.~\eqref{eq:KD-GSE}, based on mode-coupling theory like arguments, Kholodenko and Douglas have considered
explicitly a dilute suspension of colloidal hard spheres to first order in $\phi$ only, where 
$\eta_\infty$ and $\eta$ are identical, since $\Delta\eta = \mathcal{O}(\phi^2)$. For high concentrations,
we test now the validity of both the long-time and short-time versions of the KD-GSE relation, on recalling that different from $\eta_\infty$ and 
$\eta$, $d_C^S$ and $d_C^L$ are practically equal even at high concentrations. 
\begin{figure}
\begin{center}
\includegraphics[width=.35\textwidth,angle=-90]{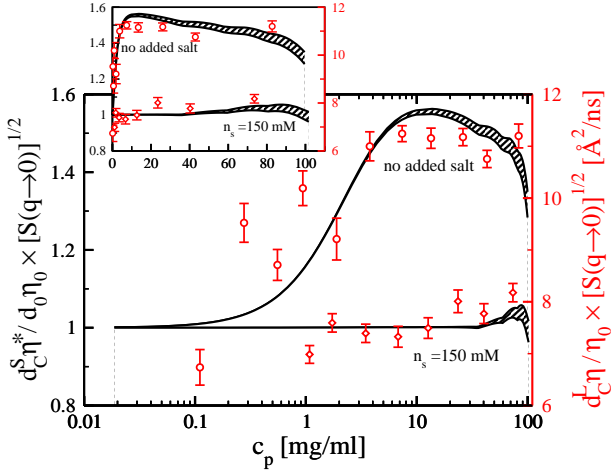}
\vspace{-2em}
\end{center}
\caption{Test of the long-time and short-time KD-GSE relations in Eq.~\eqref{eq:KD-GSE}, with $\eta^* = \eta$ and $\eta^* = \eta_\infty$, respectively.
Results for BSA solutions without added salt (upper datasets), and with $150$ mM of added NaCl (lower datasets) are shown.
Red symbols: combination of $d_C^L$ from DLS, $\eta/\eta_0$ from suspended couette rheometry, and $S(q\to0)$ from SLS.
Black lines: Theoretical results, combining $d_C^S \approx d_C^L$ and $\eta_\infty$ calculated from the 
self-part corrected $\delta\gamma$ scheme with $S(q\to0)$ from the MPB-RMSA scheme. For the long-time GSE version,
$\eta = \eta_\infty + \Delta\eta$, with $\Delta\eta$ from MCT is used.
Lower boundaries of the theoretical curves correspond to the short-time GSE, upper boundaries to the long-time version.
The theoretical curves are based on $S(q)$-input with $\phi(c_p)$ and $Z(c_p)$ concentration-interpolated using \tablename~\ref{tab:lowsalt_fitparams}.
For zero added salt, $n_s(c_p)$ was also interpolated on basis of \tablename~\ref{tab:lowsalt_fitparams}.
The parameters $L_B =0.711$ nm and $\sigma =\sigma_{B_2} =7.40$ nm are kept fixed.}
\label{fig:GSE_Test_BSA}
\end{figure}
In Ref.~\cite{Kholodenko1995}, it was argued that for uncharged hard spheres (HS) the KD-GSE relation is valid to linear order in $\phi$.
We can check this statement analytically using numerically precise 2$^\text{nd}$ order virial expansion results for $d_C^{HS} = {(d_C^S)^{HS}}$,
$\eta_\infty^{HS}$, $\eta^{HS}$ \cite{Cichocki2002,Cichocki2003,Cichocki1994}, and with $S^{HS}(q\to0,\phi)$ calculated from the precise Carnahan-Starling
equation of state. In this way, we obtain
\begin{subequations}\label{eq:GSE_test_HS}
\begin{eqnarray}
\dfrac{d_C^{HS} \eta_{\infty}^{HS}}{d_0 \eta_0} \sqrt{S^{HS}(q \to 0)} &=& 1 - 0.046\phi + 1.3713\phi^2 + \mathcal{O}(\phi^3),\nonumber\\\label{eq:GSE_test_HS_etainf}\\ 
\dfrac{d_C^{HS} \eta^{HS}}{d_0 \eta_0} \sqrt{S^{HS}(q \to 0)} &=& 1 - 0.046\phi + 2.282\phi^2 + \mathcal{O}(\phi^3).\nonumber\\\label{eq:GSE_test_HS_etastat}
\end{eqnarray}
\end{subequations}
The short- and long-time versions of the KD-GSE relation for hard spheres are identical to linear order in $\phi$, with a coefficient,
$-0.046$, which is not precisely vanishing but close to zero.
However, to quadratic order in $\phi$ already, where particle correlations come into play and $\eta_\infty$
needs to be distinguished from $\eta$, both GSE variants have distinctly non-zero virial coefficients.
Since precise values for the higher-order virial coefficients are not known to date, a test of Eq.~\eqref{eq:KD-GSE}
for larger $\phi$ can be made only using simulation and experimental data for $d_C(\phi)$, $\eta_\infty(\phi)$ and $\eta(\phi)$. 
This test has been performed in \cite{Heinen_TheoArticlePreparation}, where it is shown that both variants of the 
KD-GSE relation are approximately valid for hard spheres for $\phi \lesssim 0.1$ only.

Since neutral hard spheres are a special case of the HSY model, attained for $\gamma =0$ or $k \to \infty$, as a matter of principle the validity of the
KD-GSE relation for HSY systems is disproved already at this point. However, it still remains to be investigated in which concentration range 
the two KD-GSE relations are significantly violated
when, instead of neutral hard spheres, weakly screened, charged HSY-like particles such as charged proteins are considered.
Note here that a virial expansion cannot be reasonably applied to charged particles at lower salinity, since the pair structure functions 
and thermodynamic properties in these systems depend on $\phi$, $\gamma$ and $k$ in a non-analytical way. 

In \figurename~\ref{fig:GSE_Test_BSA}, we plot the lhs function in Eq.~\eqref{eq:KD-GSE}, both in its short- and long-time form, as a function of $c_p$.
Both BSA solutions without added salt, and solutions with $n_s =150$ mM are considered. Apart from a constant factor, which is related to the actual value
of $d_0$ in BSA solutions discussed earlier, the theoretical curves compare reasonably well to the experimental data.
There are only small differences in the short-time and long-time GSE curves in the case of BSA solutions.

With the hard-sphere-like behavior of the particles practically reached for $n_s =150$ mM, in the added-salt system the two KD-GSE variants apply
for concentrations up to $c_p \approx 50$ mg/ml, corresponding to $\phi \approx 0.1$.
For more concentrated systems, the lhs function in Eq.~\eqref{eq:KD-GSE} increases initially, going trough a shallow maximum near $c_p \approx 90$ mg/ml.
For zero added salt, violation of the KD-GSE relations is observed theoretically at all non-zero concentrations, and can be noticed in our experiment already
for $c_p \lesssim 1$ mg/ml.   
\begin{figure}
\begin{center}
\includegraphics[width=.35\textwidth,angle=-90]{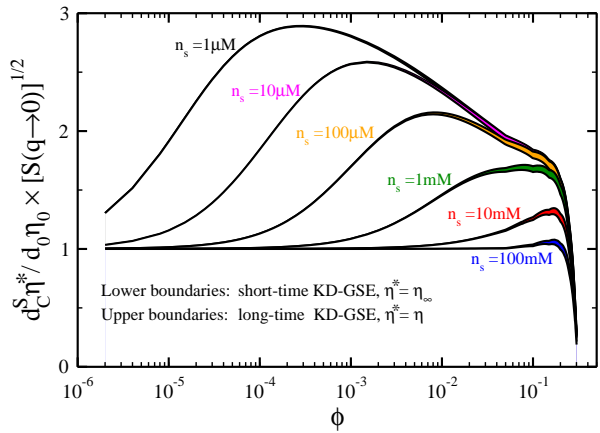}
\vspace{-2em}
\end{center}
\caption{Test of the short- and long-time versions of the KD-GSE relation in Eq.~\eqref{eq:KD-GSE} for volume fractions from very dilute values
to $\phi =30\%$, and various salt concentrations as indicated. The collective diffusion
coefficient, $d_C^{S}/d_0 \approx d_C^{L}/d_0$, and high-frequency limiting viscosity, $\eta_\infty/\eta_0$, are obtained from the
self-part corrected $\delta\gamma$ scheme. The static
viscosity, $\eta = \eta_\infty + \Delta\eta$, is calculated using MCT for $\Delta\eta$. Values for $S(q\to0)$ are obtained from the MPB-RMSA method. 
Input parameters $L_B =0.711$ nm, $\sigma =7.40$ nm, and $|Z| =10$ are kept constant.}
\label{fig:GSE_Test_various_salt}
\end{figure}

In our discussion of the KD-GSE relation, we proceed now by characterizing the crossover behavior in going from the low-salt to the high-salt regime.
To this end, in \figurename~\ref{fig:GSE_Test_various_salt}, we plot the lhs of Eq.~\eqref{eq:KD-GSE} as a function of $\phi$ for various salt contents,
using the parameters $L_B =0.711$ nm, $\sigma = 7.40$ nm, and $|Z| = 10$. These parameters are typical of aqueous solutions of small globular proteins such as BSA,
Lysozyme \cite{Goegelein2008} and Apoferritin \cite{Patkowski2005}.
The charge number $Z$ is kept constant here for simplicity. Theoretical results are plotted as a function of $\phi$ instead of $c_p$.
In lowering the salt content in \figurename~\ref{fig:GSE_Test_various_salt} stepwise by factors of $0.1$,
starting from a maximal value of $n_s =100$ mM, we find that the maximal (positive) deviation from
one of the lhs function in Eq.~\eqref{eq:KD-GSE} increases roughly logarithmically.
For low salt content, $n_s \lesssim 1$ mM, the physical origin of the maxima in \figurename~\ref{fig:GSE_Test_various_salt} is understood from 
comparing the theoretical results for $d_C$ and $\eta$ in Figs.~~\ref{fig:S0_Dcoll_Exp_Theo}
and \ref{fig:Visc_vs_conc_Exp_Theo}, respectively: The maximal violation of the KD-GSE relations occurs roughly at a volume fraction where
$d_C(\phi)$ attains its maximum, i.e. for $\phi$ determined approximately from 
$k^2_c(\phi) = k^2_s$. Recalling that $k^2_c \propto \phi$ and $k^2_s \propto n_s$, this explains why the $\phi$-location of the maxima
in \figurename~\ref{fig:GSE_Test_various_salt}
shows a power-law dependence on $n_s$ for $n_s \lesssim 1$ mM. For larger $n_s$, a crossover to hard-sphere-like behavior occurs, where
the KD-GSE relations apply for $\phi \lesssim 0.1$.


\section{Conclusions}\label{sec:conclusions}

We have investigated static and dynamic properties of aqueous BSA solutions in an integrated conceptual framework,
combining SLS/DLS, SAXS, and rheometric measurements with analytical colloid theory. Solutions with physiological concentrations of added NaCl have been studied, as
well as low-salt solutions showing distinct features in the concentration-dependence of the collective diffusion coefficient and the (reduced) viscosity.
In our analytical theoretical approach, we have used a simple spheroid-Yukawa model of BSA with isotropic, repulsive pair interactions
to calculate the static scattered intensity using the efficient MPB-RMSA method in combination with
the orientational-translational decoupling approximation. The form factor fit has been kept intentionally simple, without expecting extreme accuracy.
The resulting $S(q)$ have been used without any further fitting, in calculating
$d_C$, $\eta_\infty$, and $\eta$ on basis of our well-tested theoretical methods.
We have used the spheroid-Yukawa model for $I(q)$, and the related effective sphere-Yukawa model for the dynamic properties
as minimal models without including additional protein-specific features, to clearly reveal the pros and cons of the model.
This should help to point out more clearly the significance of left-out protein specific features.  

The measured static and dynamic properties of BSA are captured reasonably well in our simplifying SY model, with at least semi-quantitative accuracy,
for mass concentrations up to $c_p \approx 100$ mg/ml. In the range $2$ mg/ml $\lesssim c_p \lesssim 50$ mg/ml,
reliable values for the effective protein charge number, and the residual electrolyte concentration, have been obtained from the fits to the SAXS intensities.
The SAXS fits are considerably obstructed for $c_p \lesssim 2$ mg/ml by the presence of scattering impurities, and
by the breakdown of the decoupling approximation for $c_p \gtrsim 50$ mg/ml.

A well-developed maximum in the concentration dependence of the collective diffusion coefficient of BSA was found at low salinity.
This behavior is seen also in 
charge-stabilized colloidal suspensions. It is caused by the competition between electrostatic repulsion and hydrodynamic slowing down in crowded systems.
Moreover, a non-monotonic concentration dependence of the reduced viscosity of low-salinity BSA solutions was predicted theoretically,
and to some extent also seen experimentally. We have explained the local maximum in $\eta_{red}(c_p)$ as a basically non-hydrodynamic effect caused
by electric repulsion. A non-monotonic concentration-dependence of $\eta_{red}$, with a pronounced
peak at low concentration, is observed also in polyelectrolyte solutions. Thus, the low-$c_p$ peak in $\eta_{red}$ is a
generic feature of charge-stabilized dispersions at low salinity.    

An essentially concentration-independent underestimation of the experimental $d_C$ and $\eta_{red}$ by about $25\%$ and $50\%$, respectively,
is made in the theoretical predictions.
Possible reasons for this are impurity effects, and an underestimation of the corresponding single-particle coefficients $d_0$ and $[\eta]$
through our disregarding of the complex protein shape and hydration shell morphology.  

We have analyzed the
validity of a GSE relation by Kholodenko and Douglas \cite{Kholodenko1995}, which connects the collective diffusion coefficient to the shear viscosity and to the 
isothermal osmotic compressibility. Despite its appealing simplicity, the KD-GSE relation fails to capture the essential richness of macromolecular collective diffusion.
It applies to decent accuracy to electrostatically screened solutions at high salinity,
for volume fractions up to about $0.1$. However, it is violated for more crowded high-salt solutions, and for basically all volume fractions under low-salt conditions.

Possible extensions of the present work, which allow to maintain analytical simplicity to some extent, are the inclusion of short-range attractive interactions
for suspensions of larger salt content using, e.g. a two-Yukawa pair potential \cite{Chen2005, Kim2011}, and the inclusion of surface patchiness \cite{Goegelein2008}.
For the static viscosity of more strongly concentrated protein solutions than considered in the present work,
the shear stress relaxation contribution, $\Delta\eta$, can become large in comparison to $\eta_\infty$.
In calculating $\Delta\eta$, one needs then to
account for HI contributions which tend to further enlarge its value. Such an inclusion of HI effects into $\Delta\eta$ can be accomplished
on basis of an extended MCT scheme discussed in Refs.~\cite{Bergenholtz1998, Banchio1999}.  
These extensions will be the subject of a future study.

\section*{Acknowledgments}
M. Heinen is supported by the International Helmholtz Research School of
Biophysics and Soft Matter (IHRS BioSoft).
F. Zanini acknowledges a fellowship of the Institut Laue Langevin, Grenoble, France.
This work was under appropriation of funds from the Slovak grant agency VEGA 0038 for M. Antal\'ik, and VEGA 0155 and
CEX Nanofluid for D. Fedunov\'a.
F. Zhang and F. Schreiber acknowledge support from the Deutsche Forschungsgemeinschaft.
G. N\"{a}gele acknowledges support from the Deutsche Forschungsgemeinschaft (SFB-TR6, project B2).

\end{document}